\documentclass{emulateapj}
\usepackage{graphicx}
\usepackage{amssymb}
\newcommand{\psr}{PSR J0437$-$4715}

\begin{document}

\title{The spectrum of the recycled  \psr\ and its white dwarf companion}
\author{Martin Durant}
\affil{Department of Astronomy, University of Florida, FL 32611-2055, USA}
\email{martin.durant@astro.ufl.edu}
\author{Oleg Kargaltsev}
\affil{Department of Astronomy, University of Florida, FL 32611-2055, USA}
\author{George G. Pavlov}
\affil{Department of Astronomy and Astrophysics, Pennsylvania State University, PA 16802, USA\\
St.-Petersburg Polytechnical University, Polytechnicheskaya ul. 29, St. Petersburg 195257, Russia}
\author{Piotr M. Kowalski}
\affil{Helmholtz Centre Potsdam - GFZ German Research Centre for Geosciences, Telegrafenberg, 14473 Potsdam, Germany}
\author{Bettina Posselt}
\affil{Department of Astronomy and Astrophysics, Pennsylvania State University, PA 16802, USA}
\author{Marten H. van Kerkwijk}
\affil{Department of Astronomy, University of Toronto, 50 St. George Street, Toronto, Canada M5S 3H4}
\author{David L. Kaplan}
\affil{Physics Dept., U. of Wisconsin - Milwaukee, Milwaukee WI 53211, USA\\ Kavli Institute for Astrophysics and Space Research
and Department of Physics, Massachusetts Institute of Technology, Cambridge, MA 02139}

\begin{abstract}
We present extensive spectral and photometric observations of the  recycled pulsar/white-dwarf binary containing PSR J0437$-$4715, which we analyzed together with archival X-ray and gamma-ray data, to obtain the complete mid-infrared to gamma-ray spectrum. We first fit each part of the spectrum separately, and then the whole multi-wavelength spectrum. We find that the optical-infrared part of the spectrum is well fit by a cool white dwarf atmosphere model with pure hydrogen composition. The model atmosphere ($T_{\rm eff} = 3950\pm150$\,K, $\log g=6.98\pm0.15$, $R_{\rm WD}=(1.9\pm0.2)\times10^9$\,cm) fits our spectral data remarkably well for the known mass and distance ($M=0.25\pm0.02$M$_\odot$, $d=156.3\pm1.3$\,pc), yielding the white dwarf age ($\tau_{\rm WD}=6.0\pm0.5$\,Gyr). In the UV, we find a spectral shape consistent with thermal emission from the bulk of the neutron star surface, with surface temperature between $1.25\times10^5$ and $3.5\times10^5$\,K. The temprature of the thermal spectrum suggests that some heating mechanism operates throughout the life of the neutron star. The temperature distribution on the neutron star surface is non-uniform. In the X-rays, we confirm the presence of a high-energy tail which is consistent with a continuation of the cut-off power-law component ($\Gamma=1.56\pm0.01$, $E_{\rm cut}=1.1\pm0.2$\,GeV) that is seen in gamma-rays and perhaps even extends to the near-UV. 
\end{abstract}
\maketitle

\section{Introduction}
Millisecond pulsars (MSPs) are neutron stars that have been spun up through the accretion of matter from the binary companion at some time in the past, as part of a binary system. This process, known as recycling,  enables the pulsar to emit detectable emission long after it would have died as a classical pulsar.

MSPs have mainly been observed in the radio, where the emission is dominated by coherent processes, and in X-rays and $\gamma$-rays. The X-ray radiation is thought to be comprised of two components: non-thermal emission from the magnetosphere, which is characterized by hard spectra (photon index $\Gamma=1-2$) and sharp pulsations with high pulsed fraction; and thermal, surface emission, which has softer spectra and smoother pulsations. The former mechanism dominates in MSPs with high spin-down luminosity\footnote{Rate of loss of rotational kinetic energy $\dot E=I\Omega\dot\Omega$, for a standard neutron star moment of inertia $I=10^{45}{\rm\,g\,cm^2}$.} $\dot E>10^{35}$\,erg\,s$^{-1}$ (e.g., PSR B1821$-$24 and B1937+21), whereas a second group of MSPs with $\dot E=10^{33}-10^{34}$\,erg\,s$^{-1}$ (e.g., PSR J0030+0451) show mainly thermal emission, with a possible power-law (PL) high-energy component. For thermal spectra, X-ray observations are most sensitive to temperatures  $T\sim$few$\times10^6$\,K, which are considered to be typical temperatures for MSP polar caps, heated by relativistic particles.

\psr\ is an MSP ($P=5.8$\,ms) in a binary system with a white dwarf (WD), with orbital period 5.5\,d. \citet{1995ApJ...440L..81B} measured the proper motion of \psr, 135$\pm$4\,mas/yr and detected an H$_\alpha$ bow-shock nebula in the approximate direction of motion; later \cite{2008ApJ...685L..67D} greatly improved the accuracy of the proper motion measurements using VLBI ($\mu_\alpha=121.679\pm0.052$\,mas/yr, $\mu_\delta=-71.820\pm0.086$\,mas/yr), and provided the most accurate distance to \psr, $d=156.3\pm$1.3\,pc (making it one of the closest pulsars known).
The masses of the two binary components are $M_{\rm PSR}=1.76\pm0.20\,M_\odot$; $M_{\rm WD}=0.254\pm0.018\,M_\odot$ \citep{2008ApJ...679..675V}, and the orbital inclination $i=137.6\pm0.2$\degr. 
\psr\ is the brightest of the MSPs  in the UV and X-rays. Detailed spectral studies have suggested that the bulk of the X-ray emission comes from hot polar regions, where the temperature is non-uniform, decreasing with distance away from the pole (\citealt{1998A&A...329..583Z}; \citealt{2002ApJ...569..894Z}). 

For the WD companion, \citet{1993A&A...276..382D} give black-body fit of $T\sim$4000\,K to ground-based broad-band photometry. With such a cool WD, its emission no longer dominates in the  UV if the NS is hot enough. This offers a unique opportunity to observe thermal emission from the bulk of the neutron star surface in the UV/optical.

\cite{2004ApJ...602..327K} (hereafter K04) observed \psr\ for the first time in the UV with the Space Telescope Imaging Spectrometer (STIS) aboard HST. K04 found a spectrum consistent with the Rayleigh-Jeans tail of thermal emission although the measured spectral shape had a large uncertainty. Assuming the maximum area that this could be emitted from is the surface of the neutron star, they derived a surface temperature of $T\sim10^5$\,K. 

Here we describe an extensive observational campaign of\psr\ in various spectral bands. IN Section 2 we descibe the data and its analysis, one spectral window at a time. The results and spectral fits from each dataset are presented in Section 3, with a complete multi-wavelength analysis in Section 4. The results are discussed in Section 5, followed by a brief summary and outlook in Section 6.

\section{Observations and Analysis}
We utilize a suite of observations of \psr\ from ground- and space-based observatories. Here we describe in detail each part of our campaign, and the quality of the resulting data. We obtain  spectral data in the mid-IR to $\gamma$-rays, with a combination of spectroscopic and imaging photometry observations (see Table 1). We made use of ACS on HST, FORS1 on VLT/UT1 of ESO/Paranal, and PANIC on Magellan I. Furthermore, we have re-analyzed archival {\sl Chandra}, {\sl XMM-Newton}, {\sl Spitzer} and {\sl Fermi} data. The only measurements already in the literature which we have included (without reanalysis) are the ground-based photometric points of \citet{1993A&A...276..382D}, and the STIS FUV fluxes of K04. 



\begin{deluxetable}{lcccc}
\tablecaption{Log of observations of \psr\label{obs}\label{log}}
\tablewidth{0pt}
\tablehead{
\colhead{Waveband} & \colhead{Observatory/} & \colhead{Date} & \colhead{Type}  &  \colhead{Section}\\
 & \colhead{Instrument}
}
\startdata 
mid-IR & Spitzer/MIPS  &  Sep 2005 & Phot & 2.1.2\\
mid-IR & Spitzer/IRAC  & Nov 2004 & Phot & 2.1.1\\
near-IR& Magellan/PANIC & Oct 2005 & Phot & 2.2\\
optical & VLT/FORS1 &   Jul 1999 & Spec & 2.2\\
optical & HST/WFPC2 &  May 1996 & Phot & 2.4.1\\
op/UV & HST/ACS & Apr 2006 & Phot & 2.4.2\\
NUV & HST/ACS & Apr 2006 & Spec & 2.4.3\\
FUV & HST/ACS &May 2006 & Spec & 2.4.4\\
X-ray & XMM/E-MOS &  Oct 2002 & ImSpec\tablenotemark{a} & 2.5.1\\
X-ray & Chandra/ACIS &   various & ImSpec\tablenotemark{a} &2.5.2\\
Gamma & FERMI/LAT  & 2008-2010 & ImSpec\tablenotemark{a} & 2.6
\enddata
\tablenotetext{a}{imaging on a detector with intrinsic energy resolution.}
\end{deluxetable}

\subsection{Spitzer mid-IR photometry}
There are archival IRAC and MIPS {\sl Spitzer} observations of \psr.
The observations in the four {\sl Spitzer} IRAC channels, $3.6\,\mu$m, $4.5\,\mu$m, $5.8\mu$m, $8.0\,\mu$m,  \citep{Fazio2004} were carried out in November 2004, the  MIPS 24\,$\mu$m observations \citep{Rieke2004} were done in September 2005. 
The pointing accuracy of {\sl Spitzer} is reported to be better than $0\farcs5$, according to the IRAC instrument handbook\footnote{http://ssc.spitzer.caltech.edu/irac/iracinstrumenthandbook}. 
Astrometric accuracy of better than $0\farcs2$ can be reached, especially if cross-correlating with 2MASS point sources is successful. This is the case for the observations of \psr, where around 20 2MASS point sources with IRAC counterparts are in the field.
We apply the  {\sl Spitzer} Software, MOPEX (version 18.3.3)\footnote{http://ssc.spitzer.caltech.edu/postbcd/mopex.html} and its point source extraction package, APEX, for the data reduction, which we describe in detail.

\subsubsection{IRAC observations}
\begin{figure*}
\begin{center}
\includegraphics[width=\hsize]{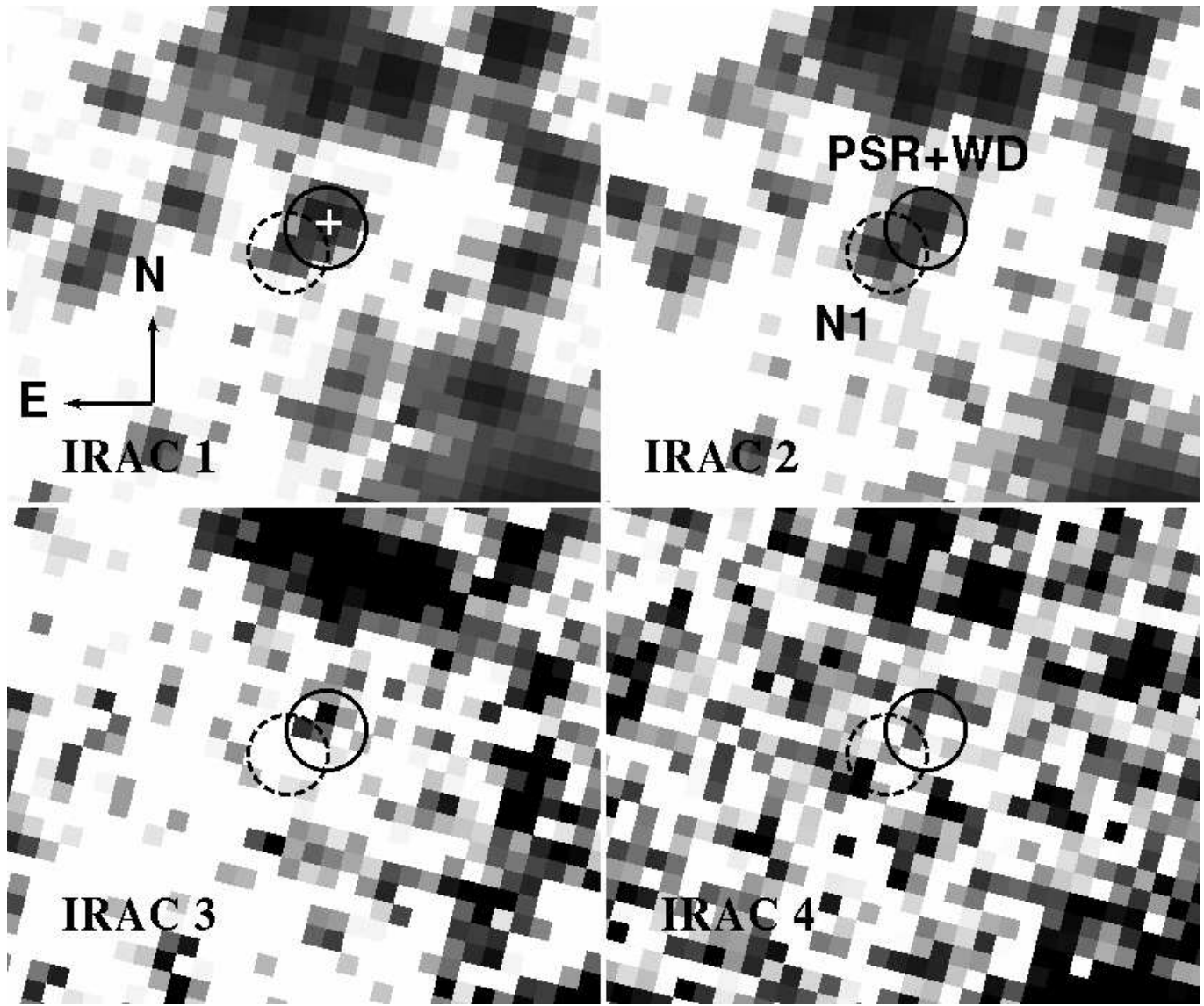}
\caption{
IRAC images of the field of J0437$-$4715 in four channels. The images are $36\arcsec \times 30\arcsec$ in size. The VLBI radio position of PSR J0437$-$4715 obtained by \citet{2008ApJ...685L..67D}, and corrected for the proper motion to the epoch of the {\sl Spitzer} IRAC observations, is marked by a white cross in IRAC channel 1. The solid  circle indicates the aperture (radius of 2 native IRAC pixels) used for the flux density measurements. 
The dashed circle corresponds to the aperture measurement done for the neighboring source N1 in channel 2. The same position and aperture was also used for channel 1 (see text), while in channels 3 and 4 N1 is not seen.
}\label{IRAC}
\end{center}
\end{figure*}

The IRAC observations are strongly affected by extended emission, which turns out to be an extended artifact in the detector, probably latent from previous observations. We select only those Corrected Basic Calibrated Data files (CBCD; they already include some pipeline processing) from the dithered observations where the target position is well separated from the artifact. This reduces the depth of the observations; in the case of the fourth channel (8\,$\mu$m), for example, only half of the available coverage can be used. For this channel we removed one more CBCD file because of a cosmic ray hit right at the target position.
The resulting mosaics at the target position are shown in Figure \ref{IRAC}; for the first channel the 
position at the epoch of the IRAC observation of \psr\ is indicated. 

{\sl Spitzer} IRAC photometric accuracy can reach $2\%$ for \emph{bright} sources, provided the proper aperture correction, the correction of the photometry array location effect, and the color correction are taken into account \citep{Reach2005}.
As recommended by the {\sl Spitzer} Science Center, we use aperture photometry instead of profile fitting photometry for the measurements, using a source aperture radius of 2 native IRAC pixels (1\farcs2 pixel size) and a sky annulus from 2 to 6 native IRAC pixels to avoid bright sources in the neighborhood. (This results in a slight over-estimation of the sky due to flux from the source in the sky aperture. For the bright background and close neighboring sources, the effect is small compared to the photometric uncertainties.)
We use the `mode' algorithm for the background estimation in the annulus to limit the influence of a faint neighbor covering part of the annulus.
The APEX source extraction clearly detects a source at the target position in IRAC channels 1 and 2, the source is marginally detected in channel 3, and not detected in channel 4. 
The following flux  values for channels 1 to 3 are aperture corrected as well as corrected for the position-dependent sensitivity,  following the IRAC data handbook recommendations: 
$16.0 \pm  0.4\,\mu$Jy;
$11.4 \pm  0.7\,\mu$Jy; and
$8.2 \pm  2.5\,\mu$Jy for channels 1, 2 and 3, respectively, where uncertainties are at the 1$\sigma$-level.
The color correction for a 4000--5000\,K blackbody is 1\% or less for each channel, so we neglect this.

The choice of the aperture and annulus size has been optimized to include as
few neighboring stars as possible.
At larger separations from the target additional constraints for background
estimates are the remaining artifacts and the varying IR background (we found
background differences of up to a factor of 3).
For IRAC channel 1 we used apertures and annuli of the same size at three
different positions in the immediate vicinity of the target to check for
conspicuous deviations in the estimated background or noise estimates. The
obtained background and noise values are comparable to those of the target
measurements. Therefore, we deduce that our aperture and annulus choice is
acceptable with no expected significant systematic effects regarding the
background or noise estimates. For the longer wavelengths, the variation in the IR background is expected to be even higher. 

In IRAC channels 1 and 2 there is a neighboring source close to the target. This source is not seen in channels 3 or 4.
The applied apertures for the target as well as for this other source are shown in Figure \ref{IRAC}; in case of channels  3 or 4 no aperture measurements of the neighbor source were required, and the dashed circles are only plotted to indicate its position.
The APEX source extraction detects the source only in channel 2; we use this position to do aperture photometry in channel 1 as well. 
The flux density of the neighbor source in channel 2 is $8.5 \pm 1.0$\,$\mu$Jy  and for channel 1 is $6.2 \pm 1.2$\,$\mu$Jy ($1\sigma$ confidence). 

Roughly 1/3 of the neighbor aperture area is also covered by the target aperture.
If the apparent neighbor were a real source, the flux of our target would be the measured value minus the neighbor's contribution; if it were instead some manner of leakage, the target flux would be the sum of both fluxes. As we don't know which of the two options takes place, we leave the flux unchanged, and increase the uncertainty by 1/3 of the apparent neighbor flux. Thus,  the target flux density in channel 2 is $11.4 \pm  3.3$\,$\mu$Jy.
Following the same argument as for channel 2,
the target flux density in channel 1 is $16.0 \pm  2.7$\,$\mu$Jy.

We note that the neighboring source is not seen in any of our other observations (X-ray to NIR). The $3\sigma$ $K_s$-band limit is 20.7\,mag (see below), indicating an emission maximum at IRAC wavelengths for this object. We suspect, therefore, that it is not a physical source\footnote{In principle, we might have detected an asteroid, but the high latitude and unusual SED \citep{2008ApJ...683L.199T} argue against this.}. 

The target is not detected in channel 4. We use the source position from IRAC channel 1 for aperture photometry in order to obtain an upper limit for channel 4. We derive a  $3 \sigma$ limit of $22.6\,\mu$Jy on the source flux.

\subsubsection{MIPS $24\,\mu$m observations}
\begin{figure}
\begin{center}
\includegraphics[width=\hsize]{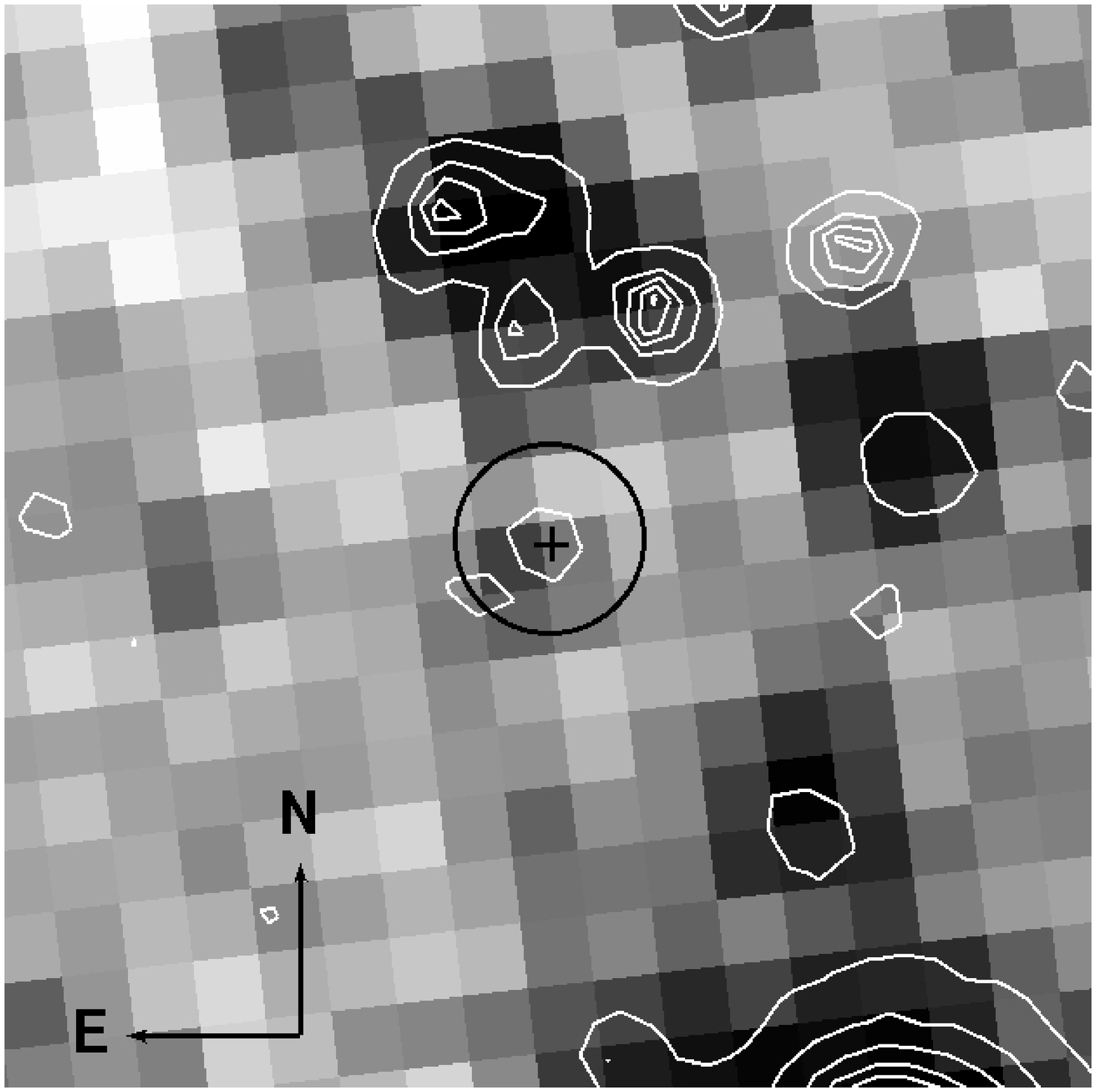}
\caption{
MIPS $24\,\mu$m mosaic for the position of \psr. The image covers 
$40\arcsec \times 40\arcsec$. The VLBI radio position of \psr\ obtained by \citet{2008ApJ...685L..67D} and corrected for the proper motion to the epoch of the {\sl Spitzer} observations is plotted as a black cross. The used aperture ($r=3\farcs5$) is also shown, together with illustrative overlay contours of the IRAC channel 2 mosaic. The small white contour south-east of the target marks the position where the faint neighbouring source is seen in IRAC 1 and 2.
}\label{MIPS}
\end{center}
\end{figure}

The photometric properties of MIPS $24\mu$m observations were reported to be excellent with achievable RMS errors of $2\%$ and an accuracy of the absolute calibration being better than $5\%$ (\citealt{Rieke2004}; MIPS instrument  handbook\footnote{ http://ssc.spitzer.caltech.edu/mips/mipsinstrumenthandbook}). 
The pipeline processed PBCD (post-BCD) $24\,\mu$m mosaic of \psr\ does not show any obvious artifact at the position of the target. 
As a check, we self-calibrated the data by dividing each EBCD (enhanced BCD) image with a normalized flat field created by taking the median of all the EBCDs. 
Self-calibration in this way removes large-scale sky variations and some artifacts. 
In our case, self-calibration did not improve the data, and we use the unchanged PBCD mosaic from the {\sl Spitzer} pipeline (version: S18.12.0) for our analysis.
At the position of the target there seems to be slightly increased count rate; however, it is close to the noise level. We use MOPEX multi-frame point response function (PRF)-fitting photometry and check our chosen parameters by producing a residual image. While we obtain a very good residual image with all prominent sources subtracted well, the source detection algorithm does not find a source at the target position even for very low detection thresholds ($2\sigma$).
Therefore we apply aperture photometry at the position of the target as derived from IRAC channel 1. The MIPS $24\,\mu$m astrometry generally agrees well with the IRAC astrometry and sources of IRAC channel 2 are over-plotted in Figure \ref{MIPS} to illustrate this.
For the aperture photometry we use a small source aperture ($3\farcs5$) and sky annulus ($6\arcsec-8\arcsec$) to avoid close-by bright neighbors. We note that the aperture correction itself has an uncertainty of around $5\%$, according to the MIPS instrument handbook. The `mode' algorithm for the background estimation in the annulus is employed. 
We derive an aperture-corrected flux density of $38 \pm 22$\,$\mu$Jy ($1 \sigma$ confidence), or $3\sigma$ upper limit of 105\,$\mu$Jy.

\subsection{Magellan NIR photometry}

We observed \psr\ with Persson's Auxiliary Nasmyth Infrared Camera
\citep[PANIC;][]{mpm+04} on 2005~October~20 with the 6.5-m Magellan~I
(Baade) telescope.  The conditions were close to photometric, with
seeing of about $0\farcs5$.  We used exposure times of $18\times 30\,$s
($K_s$-band), $9\times20\,$s ($H$-band), and $9\times 30\,$s
($J$-band), taken with small pointing dithers between the exposures.  Data reduction followed standard procedures: we masked
bad pixels, subracted dark frames and flatfielded using well-exposed
twilight flats.  Sky subtraction was done using a median-combined
image of the data.  We then combined the individual exposures
together, identified the visible stars, and masked them for a second
round of sky subtraction.  We then made a final mosaic of the individual exposures, see Figure \ref{K}.

We calibrated the photometry relative to 2MASS.  However, there was
not a sufficient number of unsaturated stars in the field of \psr\ to do this in one step.  We
therefore used other fields observed on the same night over a range of
airmasses, generally below $1.2$.  Using only stars that were still in
the linear regime of the detector and were not crowded (1--6 stars in
each field over 5 fields), we used aperture photometry in
\texttt{IRAF} with a large aperture of $1\farcs9$ radius and a sky
annulus from $2\farcs5$ to $3\farcs8$.  We corrected to airmass of 1.0
using approximate airmass coefficients for the site (0.10, 0.03, and
0.07 mag\,airmass$^{-1}$ for $J$, $H$, and $K_s$ respectively) and
determined the zero-points for the night.  The final photometry for \psr\ is $J=18.59\pm0.02$,
$H=18.37\pm0.06$, and $K_s=18.38\pm0.06$, where the errors are
statistical only. We estimate that the uncertainties in the photometric zero points were at most 0.03 mag for $K_s$ and $H$, and 0.02 for $J$.

\subsection{Optical Spectroscopy}

Spectra of the optical counterpart to \psr\ were taken on the night of
1999 July 15 using FORS1 at the Very Large Telescope.  The setup was
the same as that  used by \citet{2001A&A...378..986V} for a spectroscopic study of the faint
neutron star, RX J1856.5$-$3754 (Grism 300V with $R\approx$300, 2.6\,\AA/pix dispersion; 1\arcsec\ wide long-slit, 0.2\arcsec/pix spatial scale). The reduction followed that
described by those authors.  Briefly, two 45-minute spectra were
taken, covering the range of 3600 to 9000\,\AA\ at a resolution of
$\sim\!13$\AA, with the slit position angle chosen such that both the
counterpart and star~1 of \cite{1993ApJ...411L..83B} were covered\footnote{The star is also known as 086-010575 in UCAC3.}.  The reduction
involved bias subtraction, sky subtraction, and optimal extraction \citep{1986PASP...98..609H} of
the spectra of the counterpart using the spatial profile of star~1.

Flux calibration was done in two steps. First, we calibrated the fluxes
of star 1 using a shorter, 5-minute spectrum taken through a wide
slit (with the instrumental response determined from two flux standards;
see \citealt{2001A&A...378..986V}, especially their Sect.\ 4.4). 
Next, we estimated the slit losses by fitting a quadratic
function to the ratio of the narrow-slit to wide-slit spectra of star~1,
and used this to calibrate the fluxes of the counterpart.  Overall, we
believe our relative fluxes should be accurate to about 5\% between
4500 and 8800\,\AA.  Shortward of 4500\,\AA, the calibration is more
uncertain, since it is less clear whether our slit losses are
corrected well: the ratio of the narrow- to wide-slit spectra of star~1
shows systematic, $\sim\!10$\% deviations, likely because the wide-slit
spectra were taken at high airmass.  The absolute flux calibration is
also less certain, since there was some cirrus at the start of the
night.  We find that we have to scale the fluxes up by
8\% to match the {\em HST} photometry in \S 2.4.
 Furthermore, some regions of the spectrum are contaminated by atmospheric absorption and high background -- we have simply excluded these areas from consideration.

\subsection{HST}
We analysed several observations of \psr\ with the {\sl HST}, both imaging (WFPC2 and ACS) and spectroscopy (ACS). 
The Advanced Camera for Surveys (ACS) observations consisted originally of five visits, one in the NUV and optical using the High-resolution Camera (HRC) and the remainder in the FUV, using the Solar-blind Channel (SBC) detector. 

\subsubsection{WFPC2 Imaging}
We retrieved WFPC2 observations of the field of \psr, taken in the two wide-band filters F555W and F814W (roughly V and I band) on 1996-05-19. There were four exposures of 140\,s each, for a total of 560\,s taken through each filter. The source was located on the PC chip of the array in every case, with a pixel scale of 0.0455$\arcsec$/pix. Photometry was performed using HSTphot \citep{2000PASP..112.1383D}, a software package optimized to WFPC2 photometry by profile fitting. It efficiently rejects bad data (hot pixels and cosmic ray hits) and fits stellar profiles from libraries of pre-built PSF images (originally created using TinyTim).

For the source, we obtained (Vega-magnitudes) $m_{\rm F555W}=20.899(10)$ and $m_{\rm F814W}=19.416(8)$. Since the observations were taken early in the WFPC2 lifetime, the charge transfer efficiency (CTE) corrections are very small.

\subsubsection{ACS Imaging}

\begin{figure*}
\begin{center}
\includegraphics[width=0.45\hsize]{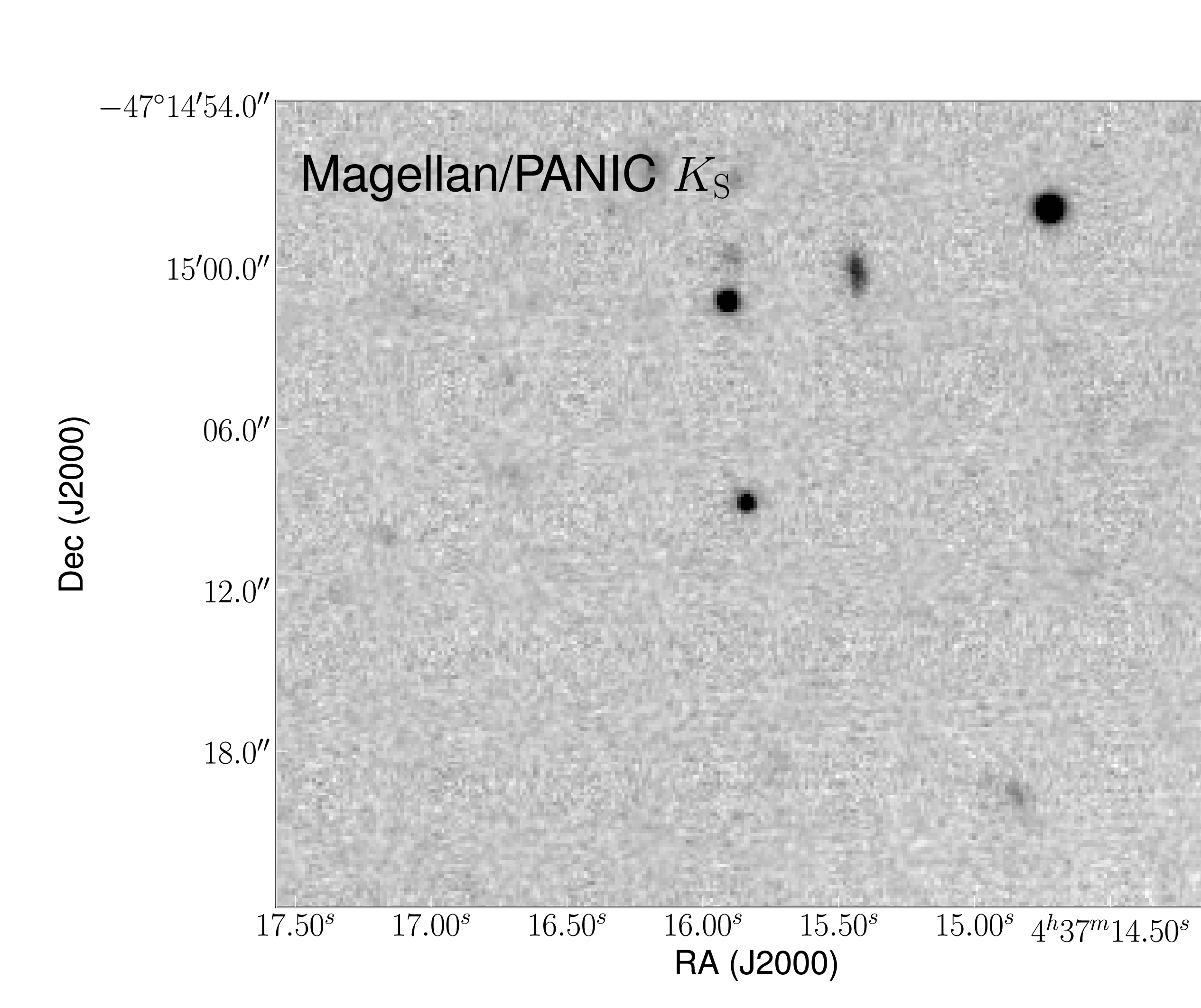}
\includegraphics[width=0.45\hsize]{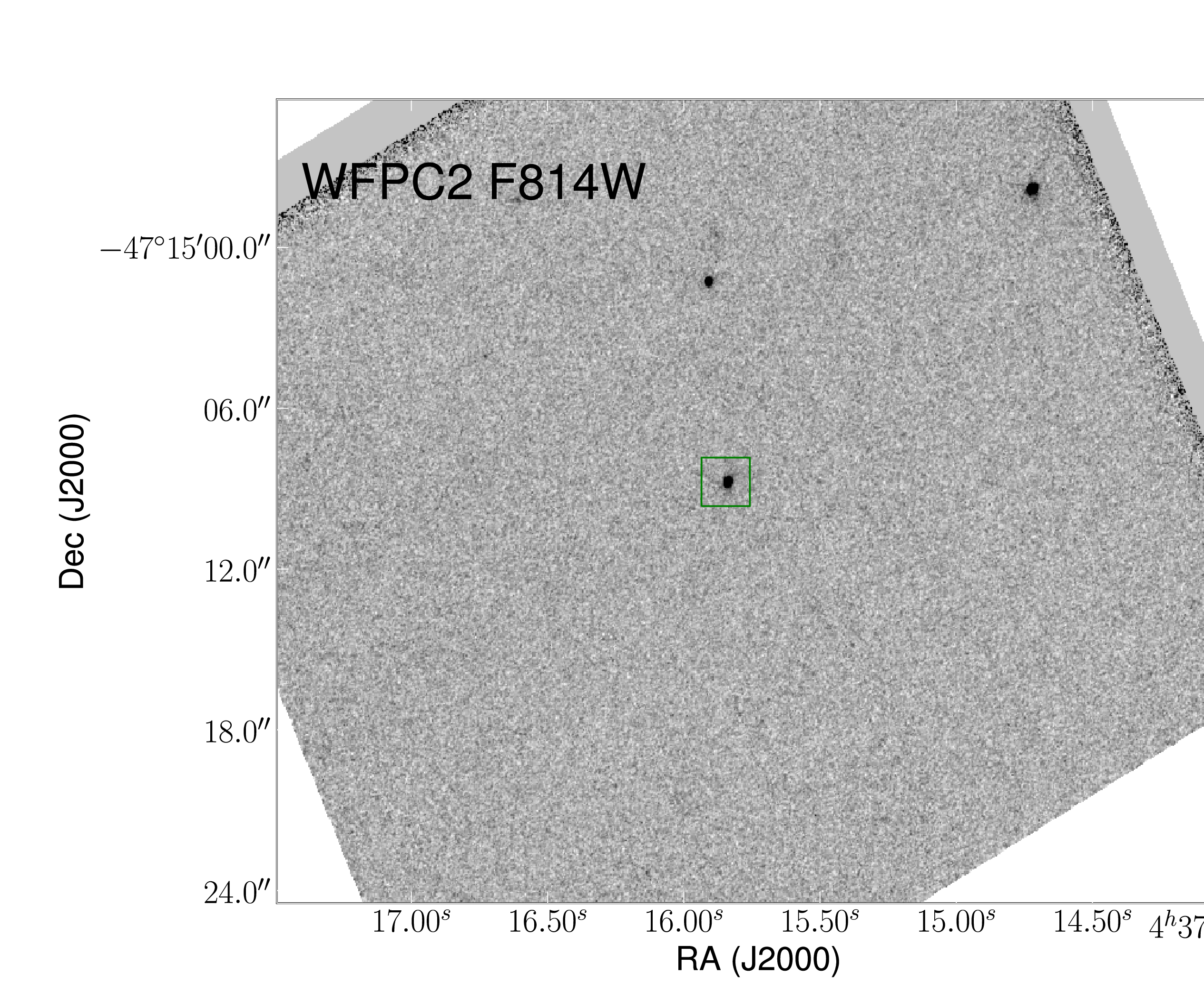}
\includegraphics[width=0.32\hsize]{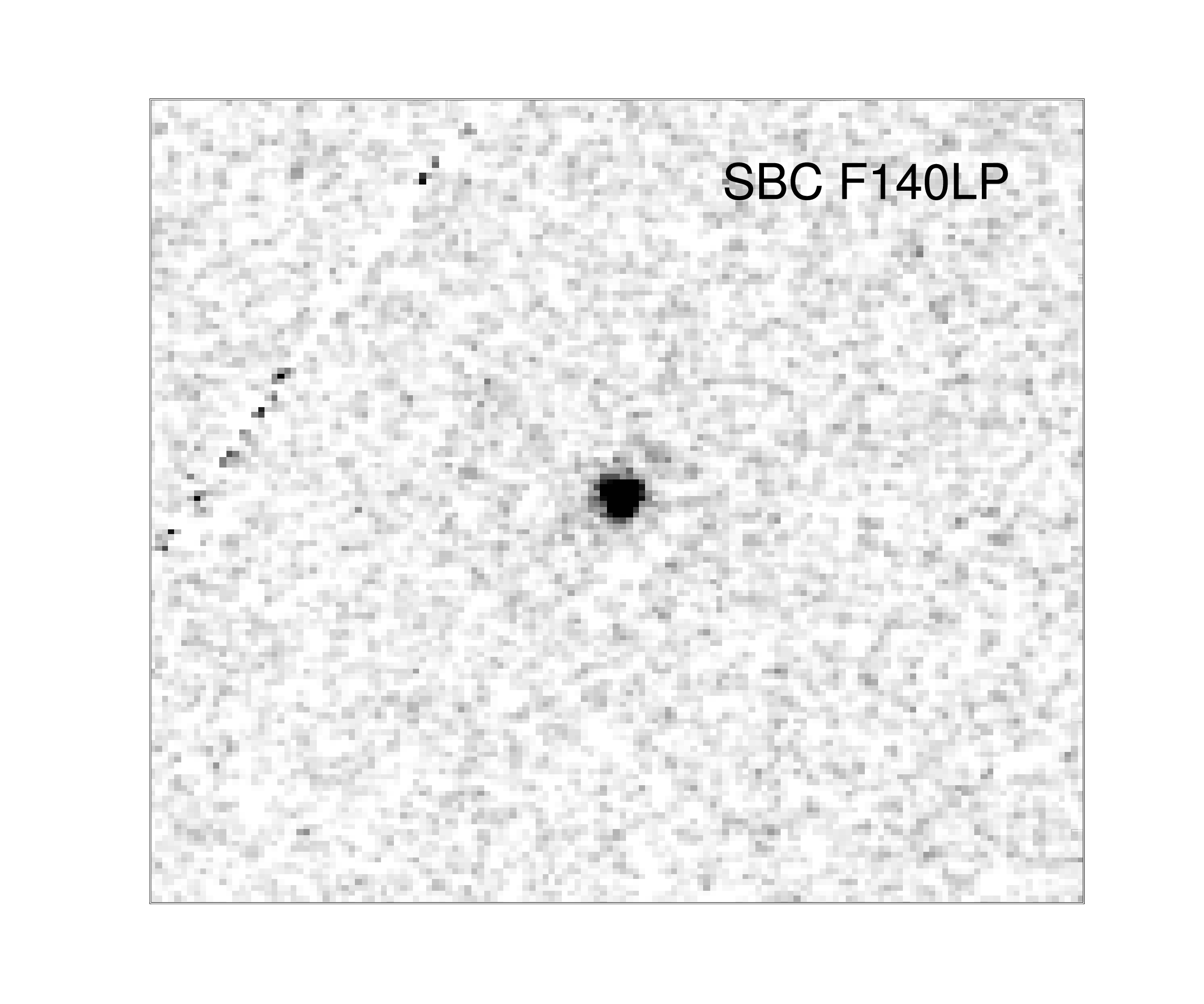}
\includegraphics[width=0.32\hsize]{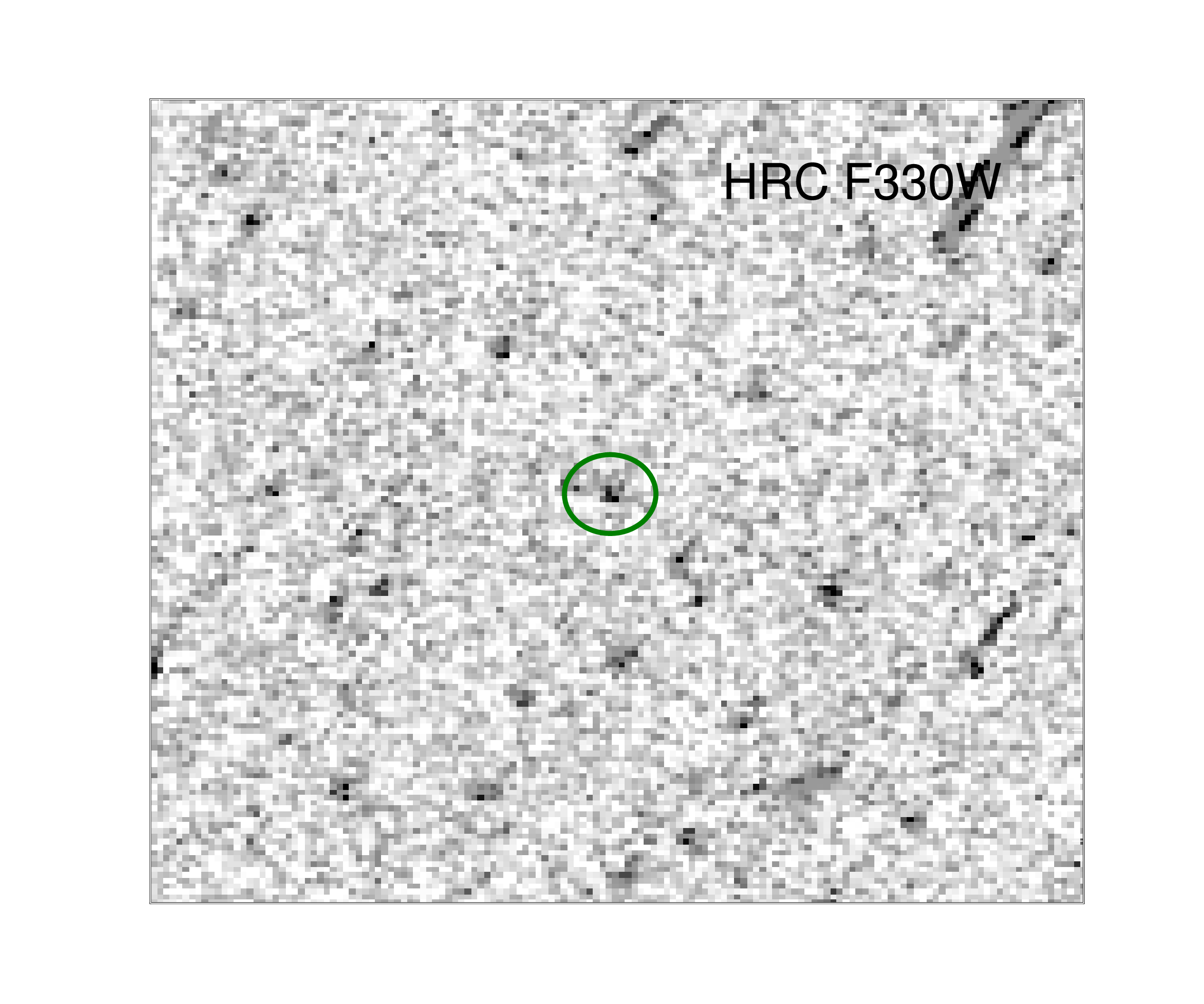}
\includegraphics[width=0.32\hsize]{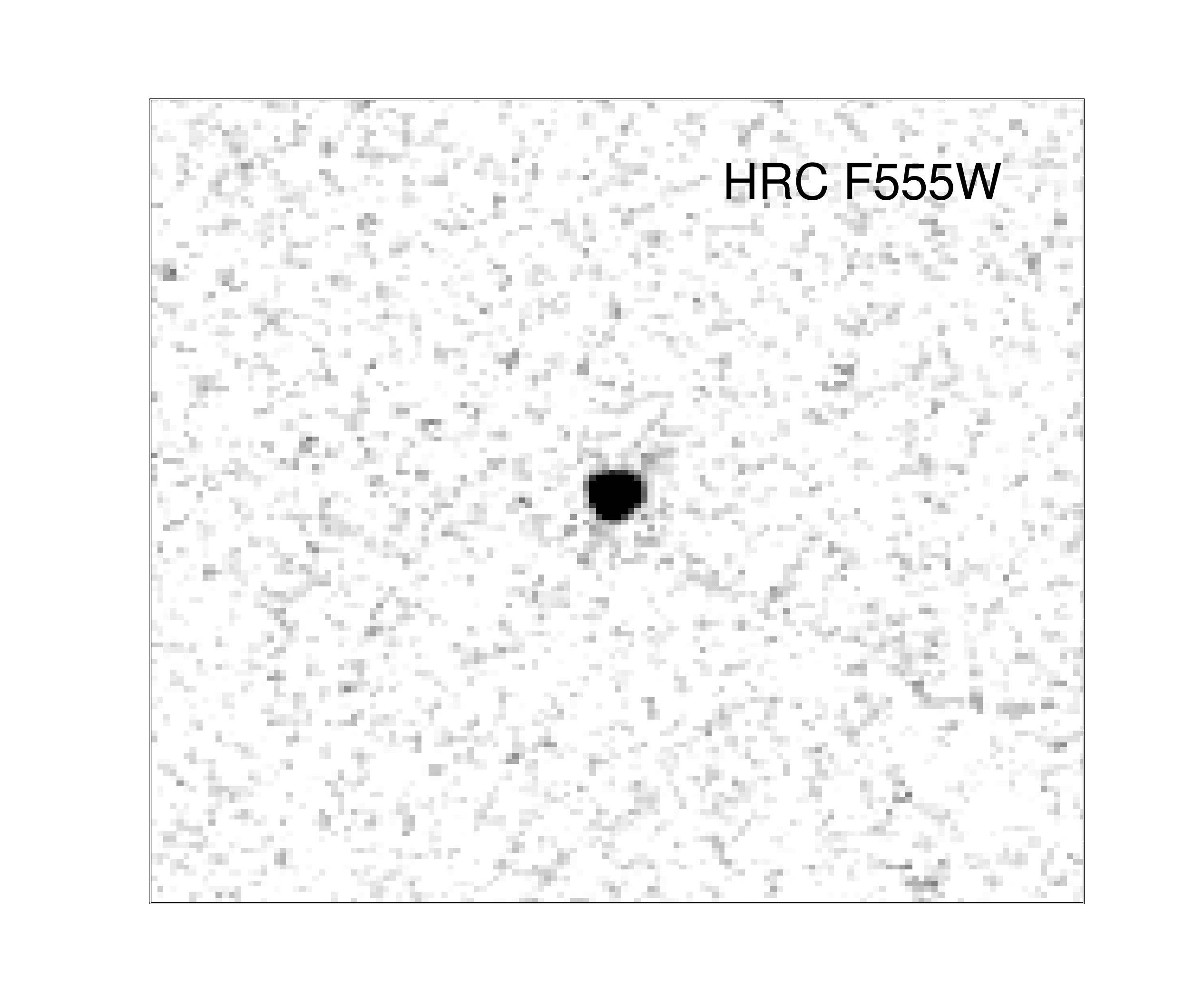}
\caption{Broad-band imaging of \psr; north is up. In each case, the source is in the center of the image. Note the cosmic rays affecting F330W (because this was a single exposure, see text; we have masked cosmic rays). No counterpart of the faint IRAC neighbor of \psr\ (Figure 1) is seen in the $K_S$-band image. The field of the ACS images is shown on the F814W image. }\label{imaging}\label{K}
\end{center}
\end{figure*}

Imaging observations are required by the prism spectroscopy, as a way to set the wavelength scale for each extracted spectrum. These images are useful, in their own right, for accurate broad-band photometry. From our observations, we have three bands, F555W, F330W and F140LP, each with a detection of the \psr\ system. One exposure was acquired in each single orbit, at the start of each orbit for the NUV and the end for the FUV. We performed aperture photometry on these, with an aperture size set to contain about 90\% of the flux, which maximizes the signal-to-noise. The correction to total flux was done using the encircled energy tables in the {\em ACS Instrument Handbook}\footnote{{\tt http://www.stsci.edu/hst/acs/documents/handbooks/ cycle18/cover.html}} (see also \citealt{2005PASP..117.1049S}), and converted to flux using the information supplied in the image headers. These are based on the latest calibration at the time of download (January 2010). Figure \ref{imaging} shows the fields centered on \psr\ in the three broad-band filters.

For F555W and F140LP, multiple images could be used to effectively remove cosmic rays and detector blemishes. We only had one F330W image to work with, however, so such rejection was not possible. No CR is seen in or around the photometry aperture.

The CTE correction for the flux in F555W is very small because the source is comparatively bright in that filter\footnote{\tt www.stsci.edu/hst/acs/documents/isrs/isr0901.pdf}. On the other hand, the F330W flux does have a substantial correction of the order 5\%-10\%. In this case, the flux is already relatively uncertain (see Table 3), and the CTE effect also increases the uncertainty.

\begin{deluxetable*}{lcccc}
\tablecaption{Broad-band photometry of \psr.\label{phot}}
\tablewidth{0pt}
\tablehead{
\colhead{Origin} &\colhead{Band} & \colhead{Central } & \colhead{Bandwidth} & \colhead{Spectral Flux $F_\lambda$ } \\
& & \colhead{Wavelength\tablenotemark{a} (\AA)} & \colhead{(\AA)} & \colhead{($10^{-18}$\,erg\,s$^{-1}$cm$^{-2}$\AA$^{-1}$)}
}
\startdata 
HST&F555W & 5356& 357 &15.58(15)\\
&F330W & 3363& 174 &1.9(2)\\
&F140LP& 1451& 127&2.93(9)\\
Magellan & J & 12200 & 2130 & 11.6(4)\\
 & H & 16300 & 3070 & 5.2(4)\\
 & K$_{\rm S}$ & 21900 & 3900 & 1.77(12)\\
Spitzer & IRAC1 & 35500 & 7500 & 0.37(6)\\
& IRAC2 & 44900 & 10150& 0.17(5)\\
& IRAC3 & 57300 & 14250& 0.07(3)\\
& IRAC4 & 78700 & 29050& $<$0.22\\
& MIPS24& 240000 & 91300& 0.38(20); $<$1.05\\
Danziger et al. &B & 4420 & 940 & 8.17(16)\\ 
&V & 5450 & 880 & 16.8(10)\\
&R & 6380 & 1380& 21.1(6)\\
&I & 7970 & 1490& 17.9(7)
\enddata
\tablecomments{ Numbers in parentheses indicate 1-sigma statistical uncertainties in the last digit(s).}
\end{deluxetable*}

\subsubsection{NUV Spectroscopy}
For the reduction of all the spectral data, we used the slitless spectroscopy package aXe\footnote{{\tt http://www.stecf.org/software/slitless\_software/axe/}}, Version 2.0 (released November 2009),  in conjunction with IRAF/PyRAF. A detailed description of the algorithms used by aXe are given in \citet{2009PASP..121...59K}. In summary, a spectral observation requires a direct (broad-band) image taken with the same pointing, which serves to calibrate the wavelength scale on the subsequent dispersed image. For the tilted spectrum, pixels within an adjustable distance in the orthogonal direction to the trace are summed. The background is determined in a region at large distance either side of the trace. 

The derivation of the wavelength solution, sensitivity curve and aperture correction are detailed in STScI Instrument Science Report (ISR) 0603\footnote{see {\tt http://www.stecf.org/documents/isr/isr0603.pdf}}.

The ACS HRC is sensitive  in the wavelength range 1700-10000\AA.
For the PR200L prism observations, we used the direct images taken in the F555W filter.  Since all the exposures for the NUV were taken during a single visit, they are very well aligned with one-another.
Although in slitless spectroscopy spectra can in principle overlap, in our case, the source is well-isolated and the brightest in the field, so such  contamination is not an issue.

Although the individual exposures are the ones used in the analysis, running {\sc multidrizzle} ensures the masking of cosmic rays as bad data, and provides an opportunity to perform visual inspection of the data. Cosmic rays are not removed, but a flag is set in the data quality (DQ) extension of the input image, marking that pixel as unusable. In order to make use of this filtering out of bad data, it is necessary to copy the modified DQ extension into the original, unmodified image file, and use this as input for aXe.

\begin{figure*}
\begin{center}
\includegraphics[width=0.75\hsize]{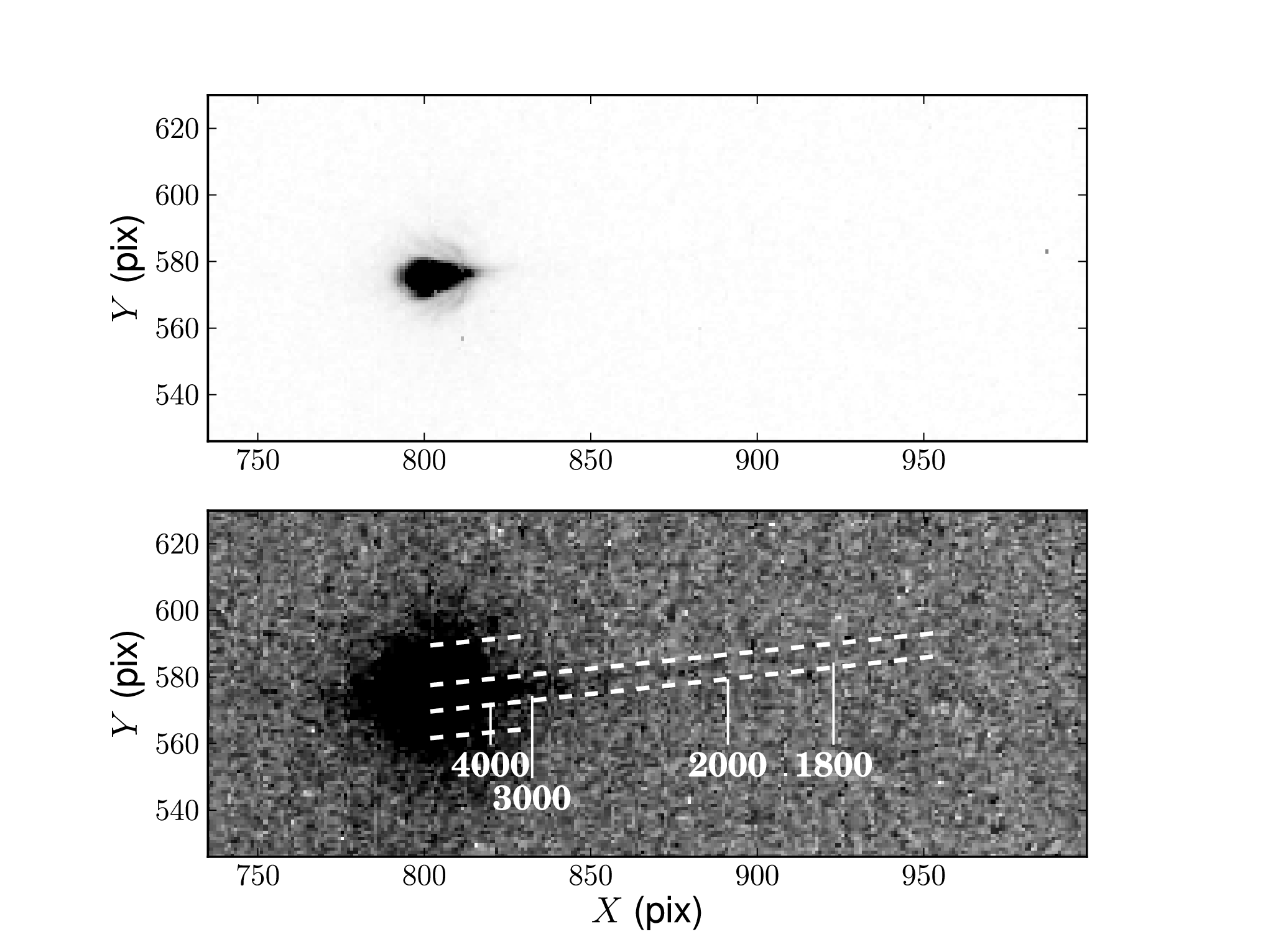}
\caption{NUV spectral image of \psr\ through prism PR200L. Wavelength decreases left-to-right, non-linearly, as marked (values in \AA). The large source seen at long wavelengths (left) is due to the {\em red pile-up}, see text. The upper panel shows the bright red source caused by the red pile up, and its surrounding reflection and diffuse structures. The lower panel is the same, but with the color scale stretched, and the extraction regions marked. The background regions used for long wavelengths are also marked; the background regions for short wavelengths lie far from the spectrum (see text).
}\label{pr200l}
\end{center}
\end{figure*}

Figure \ref{pr200l} shows the combined spectral image in the vicinity of \psr. A number of interesting features are seen. The bright elongated source is the area where the dispersion becomes very small, and all the light for wavelengths $\lambda>5000$\,\AA\ ends up centred on just a few pixels (the {\em red pile-up} effect). Since the source is intrinsically red (the cool WD dominates, see \S3.1.2), this large halo dominates over the faint spectral trace. Faint discrete structures are seen around the bright source, probably associated with reflections within the optical system, and restricted to $\lambda>4000$\,\AA. A diffuse halo of greater extent is also seen, possibly the wings of the PSF in the cross-dispersion direction. Not seen by eye, but present in the background analysis, the bright core at pixel (800,570)  also has a diffraction spike along the x-axis, at a small angle to the spectral trace. This limits the size of extraction region which will be free of contamination, particularly for intrinsically red sources (e.g., \citealt{2008ApJ...677..593W}).

We calculated final fluxes by heavily binning the count rates measured by aXe, together with the provided sensitivity curve and aperture correction (see ISR 0603\footnote{see {\tt http://www.stecf.org/documents/isr/isr0603.pdf}}).

We find, after much experimentation, that the optimum trade-off between gathering enough signal and gathering too much noise and contamination was at an aperture half-width of 1$\times$FWHM (7\,pix full width). For the background estimate we employed two regimes: at the faint, short-wavelength end ($\lambda<3175$\,\AA), where there was no bright halo, and the diffraction spike would be far from the source, we used a region from 20$\times$FWHM to 21.5$\times$FWHM from the trace. For longer wavelengths, where the halo contamination will be significant, we used a background sampled from 1.5$\times$FWHM to 3$\times$FWHM from the trace in order to subtract this contribution, but corrected the flux upwards for the small amount of source flux ($\approx$6\%) that will fall in the background sampling region. We also compared the results with those obtained using optimal extraction. In theory, optimal extraction should enhance the signal-to-noise ratio if the trace is well sampled in the spacial direction. We found that the optimal extraction produced a smoother spectrum, in better agreement with the broad-band photometric point at 3363\,\AA, so we have reason to believe that it is a fair extraction in this case; we therefore use the optimal extraction fluxes.

The final uncertainties on the fluxes are a combination of Poissonian photon noise (in both the signal and the background), small uncertainties in the sensitivity curve, and  larger systematic calibration uncertainties of 5\% in the range 1800--3000\,\AA\ rising to 15\%--20\% at the edge of the sensitivity range (K\"ummel, pers. comm.). All the uncertainties were added in quadrature. The final set of counts and fluxes for the wavelength bins chosen are listed in Table \ref{nuv}.

\begin{deluxetable*}{cccccc}
\tablecaption{FUV and NUV spectral data\label{nuv}\label{fuv}}
\tablewidth{0pt}
\tablehead{
\colhead{Wavelength bin} & \colhead{Bin width} & \colhead{Gross Counts} & \colhead{Background Counts} & \colhead{Net Counts\tablenotemark{a}}& \colhead{Flux\tablenotemark{a} $F_\lambda$} \\
\colhead{centre (\AA)} & \colhead{(\AA)} &&&& \colhead{($10^{-18}$\,erg\,s$^{-1}$cm$^{-2}$\AA$^{-1}$)}
}
\startdata 
1245 & 30 & 403 & 225 & 178(20) & 6.42(0.80)\\
1275 & 30 & 424 & 177 & 247(21) & 5.34(0.53)\\
1305 & 30 & 412 & 144 & 268(20) & 5.19(0.48)\\
1335 & 30 & 387 & 106 & 281(20) & 5.36(0.47)\\
1365 & 30 & 348 & 93 & 255(19) & 5.00(0.45)\\
1395 & 30 & 256 & 70 & 186(16) & 3.86(0.39)\\
1425 & 30 & 253 & 70 & 183(16) & 4.05(0.41)\\
1455 & 30 & 171 & 51 & 120(13) & 2.86(0.35)\\
1485 & 30 & 188 & 49 & 139(14) & 3.60(0.40)\\
1515 & 30 & 195 & 49 & 146(14) & 4.16(0.46)\\
1545 & 30 & 132 & 39 & 93(11) & 2.99(0.40)\\
1575 & 30 & 128 & 41 & 87(11) & 3.18(0.45)\\
1605 & 30 & 119 & 39 & 80(11) & 3.36(0.50)\\
1645 & 50 & 163 & 55 & 108(13) & 3.27(0.43)\\
1695 & 50 & 107 & 41 & 66(10) & 2.42(0.41)\\
1745 & 50 & 111 & 41 & 70(12) & 3.36(0.55)\\
1795 & 50 & 99 & 41 & 58(10) & 4.26(0.78)\\
1845 & 50 & 53 & 27 & 25(7) & 3.16(0.94)\\
\hline\vspace{-0.05in}\\
2150 & 700 & 54808 & 52685 & 2123(230) & 1.54(0.64)\\
2600 & 200 & 7699 & 7006 & 693(88) & 1.07(0.15)\\
2750 & 100 & 3053 & 2648 & 405(55) & 1.27(0.18)\\
2850 & 100 & 3127 & 2731 & 396(56) & 1.15(0.17)\\
2950 & 100 & 3167 & 2516 & 651(56) & 1.72(0.17)\\
3050 & 100 & 2177 & 1689 & 488(47) & 1.17(0.13)\\
3150 & 100 & 2620 & 1894 & 726(51) & 1.68(0.15)\\
3300 & 200 & 4635 & 3243 & 1392(68)& 1.73(0.27)\\
3500 & 200 & 4096 & 2377 & 1719(64)& 2.33(0.36)\\
3700 & 200 & 2506 & 1203 & 1303(50)& 1.92(0.30)\\
3900 & 200 & 2824 & 1238 & 1586(50)& 2.54(0.41)
\enddata
\tablenotetext{a}{Uncertainties are statistical only for the net counts, and include both statistical and systematic/calibration contributions for the flux. Numbers in parentheses show one-sigma uncertainties.}
\end{deluxetable*}

\subsubsection{FUV Spectroscopy}
\begin{figure*}
\begin{center}
\includegraphics[width=0.75\hsize]{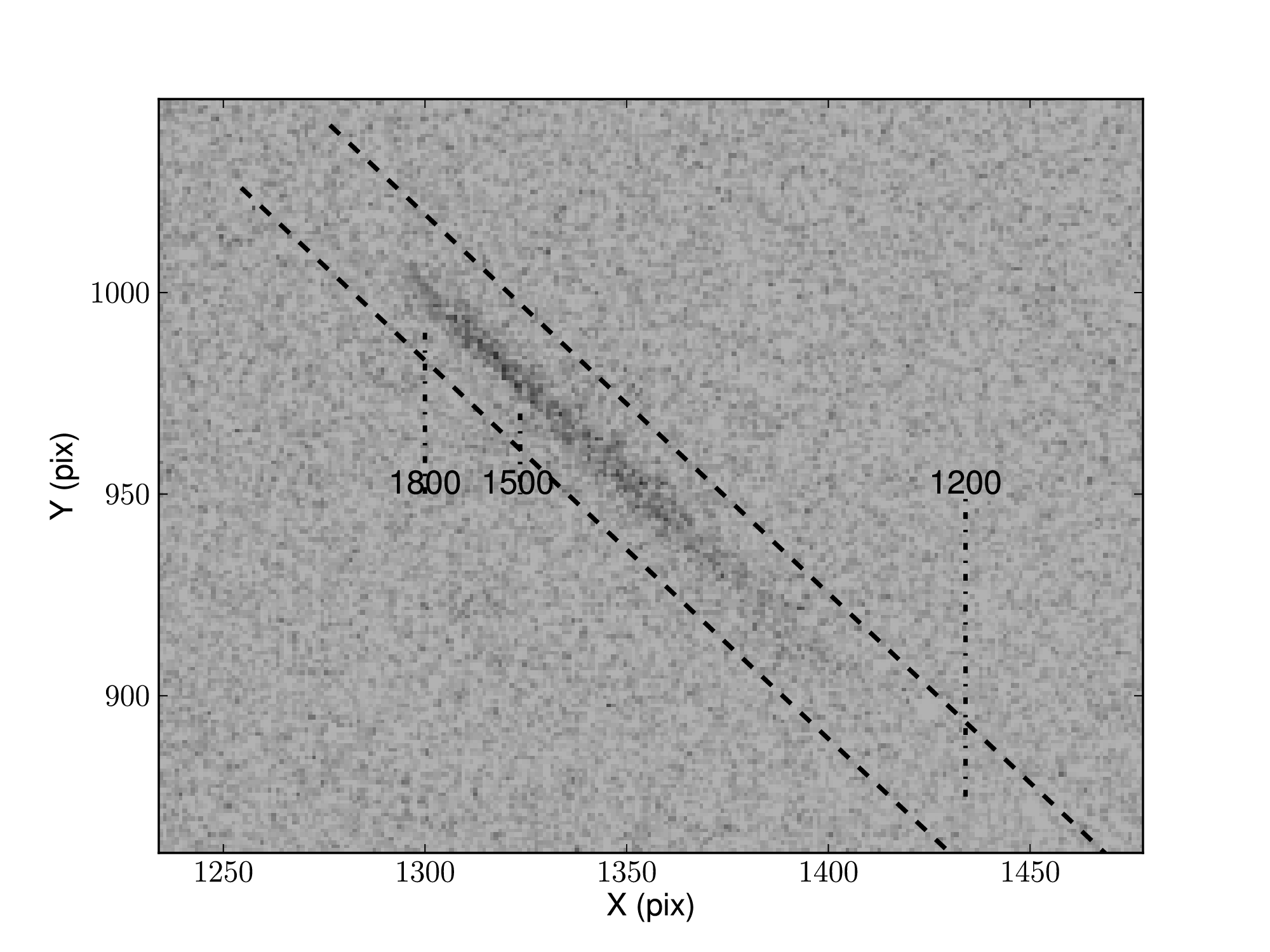}
\caption{FUV combined spectral image of \psr\ through prism PR130L, as produced by {\tt multidrizzle}. The extraction region is labeled, along with approximate wavelengths in \AA.}\label{pr140l}
\end{center}
\end{figure*}

\psr\ was imaged through the PR130L prism. This prism was chosen in order to decrease the background level, since it has low throughput shortward  of 130\,nm, and thus limits the contribution of the brightest geocoronal  lines. Furthermore, the spectral exposures were taken while the spacecraft was in the Earth's shadow, further reducing the background. The imaging exposures were done in the remainder of the orbit. In addition, the SBC's Multi-Anode Microchannel Array (MAMA) detector does not suffer from cosmic ray hits as do CCD detectors. As the resulting combined image in Figure \ref{pr140l} shows, the background is faint and uniform, and the spectrum is very easy to see. In comparison to the NUV observations, the spectral trace is well-defined, and there is no background structure or any halo/pile-up effects.

Reduction proceeded in a similar way to the NUV spectral analysis above. Since the exposures come from separate visits in this case, and therefore required re-acquisition of a guide star, alignment between them is not guaranteed, and the source position on the direct image was translated back to each input direct image using the aXe task {\sc iolprep}. This was particularly important for the Visit 02 exposures, because of the loss of the guide star during this visit. The data had an enhanced background level, and the data from the last orbit were totally lost. Nevertheless, we do include the data from the first part of the visit, and we find that it improves rather than decreases the signal-to-noise ratio. Before proceeding with the extraction, the input coordinates need to be corrected for an offset caused by the choice of broad-band filter for imaging (see ISR 0602\footnote{\tt http://www.stecf.org/documents/isr/isr0602.pdf}).

Due to the much lower background level compared to the NUV, and absence of the problems associated with red pile-up and diffraction spikes near the spectral trace, it proved to be optimal to extract the source from a somewhat wider aperture, half-width of 4$\times$FWHM (19\,pix full width). See Figure \ref{pr140l} for the locations of the extraction region.
The results are listed in Table \ref{fuv}. With the low sensitivity, the systematic uncertainties in the calibration make less of an impact in this case, and optimal extraction produced a spectrum that looks a little noisier, but otherwise similar - we therefore used simple fixed aperture extraction.  Systematic uncertainties are of the order 5\% throughout the spectral band analyzed.

\subsection{X-rays}
We retrieved one long {\sl XMM-Newton} observation and nine {\sl Chandra} observations from the archives, and re-processed all the data to be analyzed simultaneously. The data consist of CCD imaging and provide low-resolution spectra using the inherent energy resolution of the detector, as opposed to inserting a spectral element such as a grating. Although the Reflection Grating Spectrometer (RGS) instrument of {\sl XMM-Newton} does provide such data, its count rate was too low for consideration. 

\subsubsection{XMM-Newton}
{\sl XMM-Newton} observed  \psr\ on 2002-10-09, 03:32:34--22:50:04. This observation and its analysis are described in detail by \citet{2006ApJ...638..951Z}. Here we consider the data from the two EPIC-MOS detectors, operated in Full Window mode (2.6\,s time resolution) and the EPIC-PN detector operated in Fast Timing mode (0.03\,ms time resolution). The raw data were processed into calibrated event files using the standard tasks of the provided software XMM-SAS, version 9.0 (20090615\_1801). After filtering the data for times of flaring, the total live time was 63.0\,ks for the MOSs and 62.9\,ks for the PN.


For the MOS data sets, we extracted the source spectrum from a circle of 15\arcsec\ radius centered on the source. This only captures some 70\% of the total flux, but such a compromise is necessary because of the relatively high background and the background from a circular annulus with inner and outer radii of 114\arcsec and 200\arcsec, respectively. 

In timing mode, all information about the Y-axis position of a photon on the detector is lost. We filtered out low-energy electronic noise\footnote{{\tt http://www.star.le.ac.uk/$\sim$amr30/BG/BGTable.html}, column titled Electronic Noise} by keeping only energies $E>600$\,eV. 
There is a bright AGN in the field whose one-dimensional PSF partly overlaps with that of the pulsar (see \citealt{2002ApJ...569..894Z}). We therefore extracted counts from an aperture chosen to avoid contamination, that only contained 61\%--66\% of the flux (higher fraction for lower energies).  We correct the spectrum for this finite-aperture effect.

\subsubsection{Chandra}
We retrieved nine {\sl Chandra} ACIS imaging datasets, with IDs 6154, 6157, 6767, 6768, 741, for a total exposure time of 73.2\,ks. We do not consider High Resolution Camera (HRC) observations nor ACIS observations affected by pile-up. The observations have a range of exposure times and off-axis distances. The total exposure time was 73.2\,ks.

The source extraction region was chosen to be fairly large and uniform for all the observations, as the background is low, and the varying off-axis distance for each observation means the size of the PSF is different. The radius was set at 9.15\,pix (4\farcs5), which includes some 99\% of the flux for the on-axis PSF, and most ($\approx$90\%) even for the broadest PSF in the set of observation. The background was taken either from an annular region centered on the source, or rectangular regions either side of the readout direction, for observations with subarrays, where the active area is restricted in one direction.

%

\subsection{Fermi}

We  retrieved  data on the region containing \psr\ collected in 18 months of {\sl Fermi} LAT observation.  $\gamma$-ray data has much lower count-rates than X-rays, extended, broad-winged PSF and frequency-dependent energy resolution. Standard filtering (see \citealt{2010ApJS..187..460A}) was applied. The reduction to obtain a spectral model or fluxes for a single source is therefore a likelihood maximization method, with explicit consideration of nearby sources and their respective spectra. Since spectral fitting is then an intrinsic part of the extraction process, we describe it in Section \ref{gam}.

\section{Spectrum and fits}

\begin{figure*}
\begin{center}
\includegraphics[width=0.9\hsize]{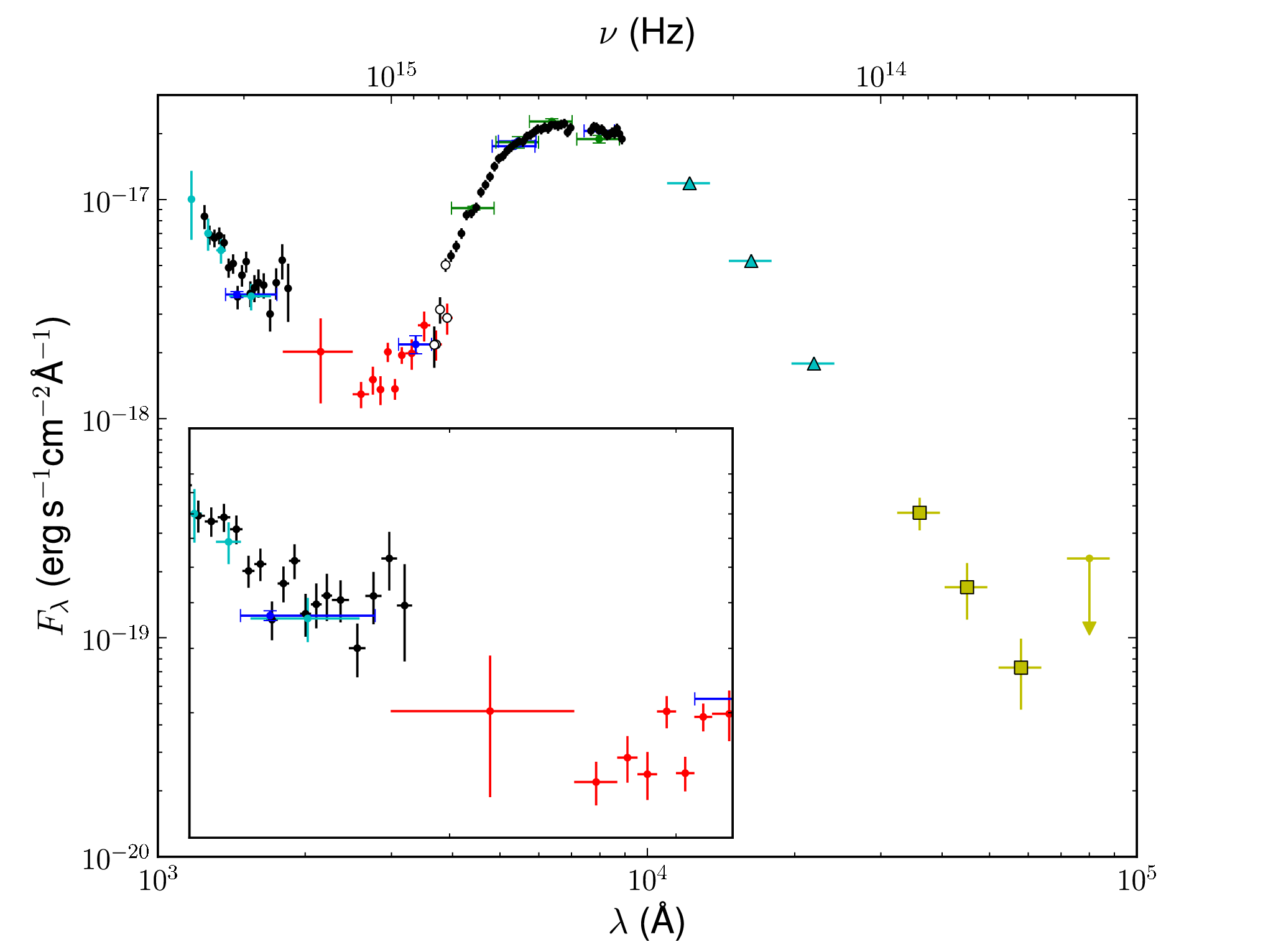}
\caption{UV to mid-IR spectrum of \psr\ with a zoom of the UV portion in the inset. Black points are from prism PR130L (UV) and VLT/FORS1 (optical), red points from prism PR200L, blue points from {\sl HST} broad-band imaging, cyan triangles from Magellan/PANIC IR photometry, and yellow squares (and a limit) from {\sl Spitzer} photometry. Cyan UV points are from previous STIS spectroscopy (K04), and green points are ground-based photometry from \citep{1993A&A...276..382D}. Open markers denote fluxes from FORS1 or PR200L with uncertain calibration. }\label{spec}
\end{center}
\end{figure*}

In Figure \ref{spec} we show the UV-IR spectrum of J0437. Note that we don't include the  MIPS point, since it is more likely to be a (non-restrictive) upper limit.
It is reassuring to see the high degree of agreement between our FUV spectral points, the F130LP broad-band photometry  and the previous STIS spectrum. Likewise, the NUV spectrum and the F330W photometry  agree, but there is no independent data in this region to verify the spectral shape. Furthermore, the NUV and optical points support a smooth continuum linking the two spectral regions. Given the several problems inherent in processing the PR200L data, we regard with suspicion the hint of structure around 3700\,\AA.

The extinction to the source is not well constrained by fits to the X-ray spectrum, but must be small, with the hydrogen column of $\sim10^{20}$\,cm$^{-2}$ or less \citep{2002ApJ...569..894Z}. This is equivalent to $E(B-V)<0.05$, consistent with the small distance to the source, $d=156.3$\,pc. We therefore attempt to model our combined spectrum with extinction  values $E(B-V)=0.01$, 0.03 and 0.05. Furthermore, there is some considerable difference in the extinction curves in the literature, for a given value of extinction. This is an additional uncertainty in our fits. In practice, the shapes of the extinction curves in, for instance, \citet{1979MNRAS.187..785S} and \citet{1989ApJ...345..245C} are very similar, but different in normalization. We will use \citet{1989ApJ...345..245C}'s more recent parameterization  for our fitting. 

In the NUV and FUV we considered only the spectroscopic fluxes because of the lower signal-to-noise of the F330W point and the extreme asymmetry of the F130LP band-pass. 

\subsection{UV to mid-IR}
\subsubsection{Analytic models}
\begin{figure*}
\begin{center}
\includegraphics[width=0.8\hsize]{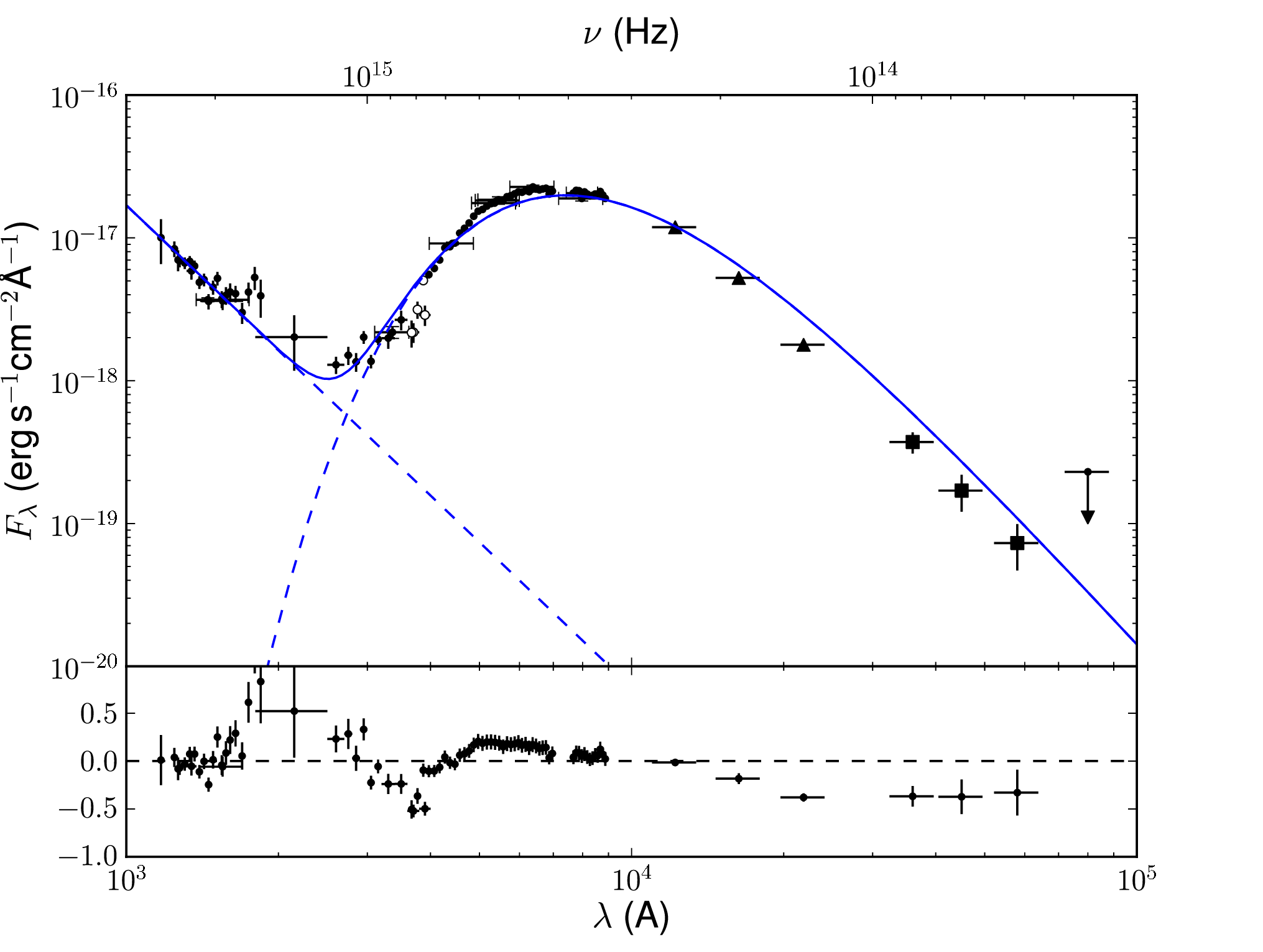}
\includegraphics[width=0.8\hsize]{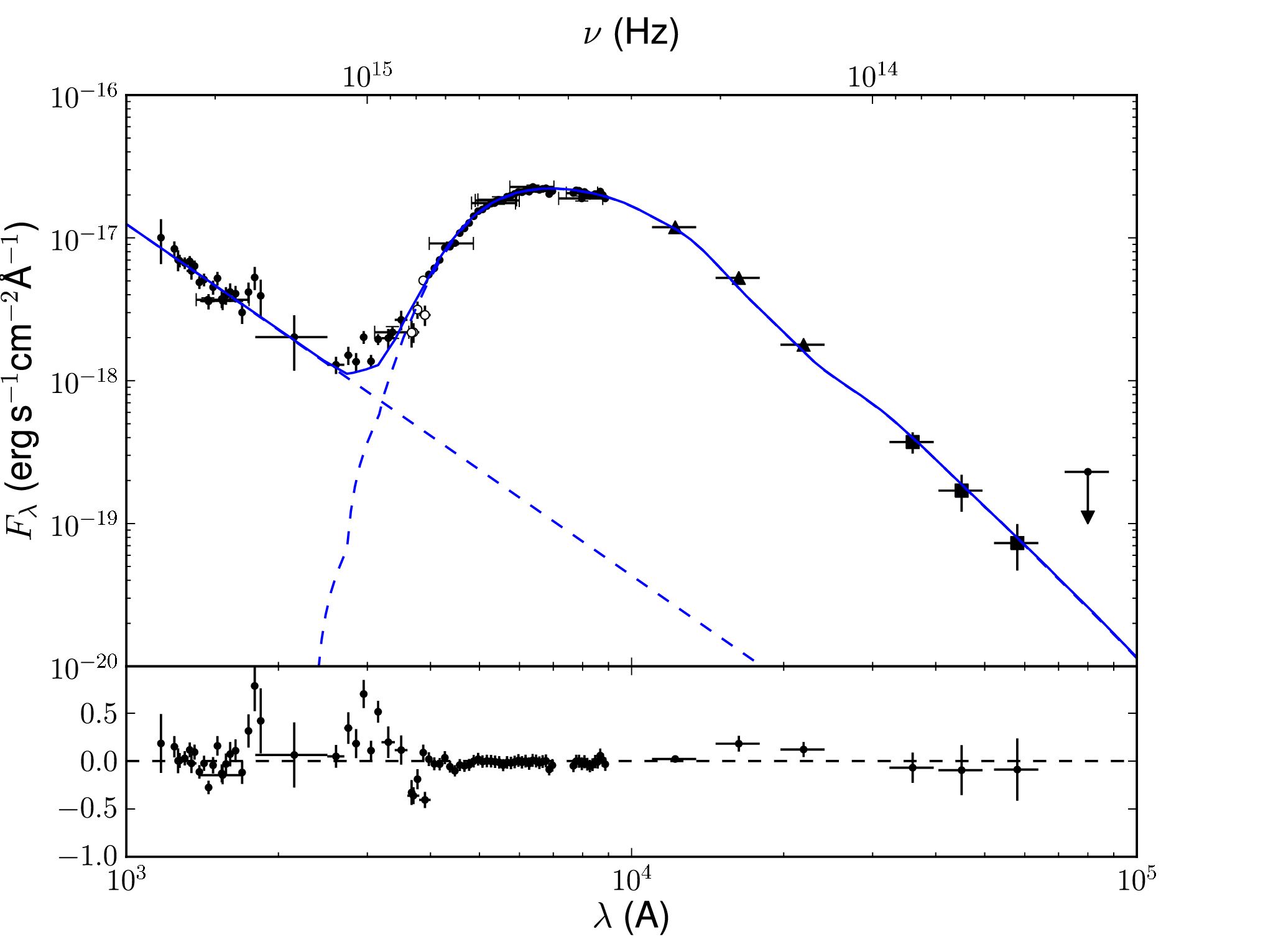}
\caption{{\em Top:} Black-body plus power-law fit to the spectral data for \psr. This is the case for the preferred value of reddening $E(B-V)=0.03$, giving $T_{\rm BB}=3910$\,K, $\alpha=-3.4$. The lower panel shows fractional residuals  (flux-model)/model.
{\em Bottom:} Model with white dwarf atmosphere ($T_{\rm eff}=3950$\,K, $\log g=6.98$, pure H) and power-law, with index $\alpha=-2.46$ (solid blue). The two components are shown separately (dashed blue). The lower panels show fractional residuals:  (flux-model)/model.}\label{analyt}\label{WD}\label{WD_PL}
\end{center}
\end{figure*}

We first fit the observed spectrum with simple analytical models. In particular, the maximum in the optical can be fitted with a black-body (BB) curve, and the FUV-NUV with a power-law ($F_\lambda\propto\lambda^\alpha$; PL). We attempt this for different reasonable values of reddening, $E(B-V)=$\,0.01, 0.03, 0.05. The results of these fits are shown in Table \ref{bb_pl}. It should be noted that none of these fits is statistically acceptable (see Figure \ref{analyt}), but the numbers can be taken as indicative. The spectrum shortward of 1500\,\AA\ appears to be well-represented by the $\alpha\approx-3.5$ power-law.

We can see that a simplistic sum of a $\sim$4000\,K black-body and a PL with $\alpha=-(3-4)$ (i.e., close to a Rayleigh-Jeans tail of a thermal spectrum with a high temperature) roughly describes the data. The apparent BB emitting radius is about 20,000\,km (or 0.028\,$R_\odot$).

\begin{deluxetable}{ccccc}
\tablecaption{Fits to IR-UV spectrum\label{bb_pl}}
\tablewidth{0pt}
\tablehead{
\colhead{$E(B-V)$} & \colhead{$T$ (K)} & \colhead{$R$\tablenotemark{a}} & \colhead{$\alpha$}  &  \colhead{$N_{\rm PL}$\tablenotemark{b}}
}
\startdata 
&\multicolumn{3}{c}{BB+PL}\\
0.01 & 3870(40) &  2.02(11) & $-$3.3(4) &  3.7(2) \\
0.03 & 3910(40) &  1.97(11) & $-$3.4(4) &  4.3(3) \\
0.05 & 3950(40) &  1.95(11) & $-$3.5(4) &  5.0(3) \\
\\
&\multicolumn{3}{c}{WDA+PL}\\
0.03 & 3950(150)& 1.87(20) & $-$2.46(17) & 4.61(19)
\enddata
\tablecomments{Black-body plus power-law fits (\S 3.1.1) and WD atmosphere model plus power-law fit (\S 3.1.2). Uncertainties are 1-sigma, taken from the diagonal of the covariance matrix and assuming the fit is reasonable.}
\tablenotetext{a}{Effective radius for a distance of 156.3\,pc, units of 10$^4$\,km.}
\tablenotetext{b}{Spectral flux of the PL component at 1500\,\AA, in $10^{-18}$\,erg\,s$^{-1}$cm$^{-2}$\AA$^{-1}$.}
\end{deluxetable}

\subsubsection{WD atmosphere}\label{WDat}
Although informative, the BB only crudely approximates the optical-IR spectrum above and cannot
reveal information on the surface properties of the WD. A better description could be provided by a detailed white dwarf (WD) atmosphere model. Atmosphere models have been developed for a variety of WD masses, ages and compositions (e.g., \citealt{1995ApJ...443..764B}). Note that, in principle, the WD surface is irradiated by the higher-energy emission of the companion pulsar, but the effect is small: for the known binary separation  \citep{2008ApJ...679..675V}, the radiative intensity from the pulsar  (see also \S4) at the distance of the WD is approximately 3$\times10^8$\,erg\,s$^{-1}$cm$^{-2}$, about 2\% of the outward radiative flux; we therefore neglect it in our modeling. We describe in detail the model fitting process. 

The spectrum of the WD companion was modeled using  white dwarf atmosphere (WDA) models that account for recently introduced improvements in the description of the atmosphere of these stars, especially Ly-$\alpha$ absorption in the UV \citep{2006ApJ...651L.137K}, opacity of helium dominated medium \citep{2006ApJ...641..488K,2007PhRvB..76g5112K}
and refraction \citep{2004ApJ...607..970K}. The models used in the analysis were successful in describing the entire spectral energy distributions of many cool WDs,
including the UV spectrum of pure-H atmosphere WD BPM 4729 \citep{2006ApJ...651L.137K} and  the near UV spectra of WDs with $T_{\rm eff}\sim4000 \, K$ \citep{2010ApJ...715L..21K}.

The fit in effective temperature and gravity for  He and H atmospheres was performed for the known photometry, BVRI of \citet{1993A&A...276..382D}, and  Magellan JHK
and {\sl Spitzer} IR measurements reported here, assuming that the distance to the system is $156.3$\,pc \citep{2008ApJ...685L..67D} and the reddening is $E(B-V)=0.03$.
 The  spectral energy distribution of the WD is best reproduced by a
pure hydrogen atmosphere model. The hydrogen is clearly visible, when comparing with pure-He model spectrum, by strong absorption in the UV, due to strong Ly-$\alpha$ red wing
opacity \citep{2006ApJ...651L.137K} and flux suppression in the IR due to collision-induced absorption (CIA)  by molecular hydrogen \citep{1999ApJ...511L.107S}. The best-fit model closely matches the VLT/FORS1 spectrum (see Figure \ref{WD_PL}).

The best-fit  effective temperature, $T= 3950\pm150$\,K, is in agreement with previous estimates \citep{1993Natur.364..603B,2006A&A...450..295B,1993A&A...276..382D} and the BB fit from above, see Figure \ref{WD}. The obtained gravity ($\log g=6.98\pm0.15$ [cgs]) fits the models
of a helium core white dwarf \citep{2001AN....322..405S,2009A&A...502..207A}, assuming the mass of the companion to be 0.254\,$M_{\sun}$ \citep{2008ApJ...679..675V}. The uncertainties are estimated from the uncertainty in the photometric fluxes (Table \ref{phot}), and the small uncertainties in the distance and WD mass. 
We include the fit parameters in Table \ref{bb_pl}; we also list the radius, $R$, which is fitted
implicitly via $\log g$ (because the mass is well known); the quoted error reflects uncertainties in the
temperature and distance. Without spectral lines, the spectrum
provides no distance-independent constraints on the gravity).


While some of cool WD pulsar companions have optical photometry,
this is usually not enough for definitive assignment of their atmospheric chemical composition and parameters
(see discussion in \citealt{2006A&A...450..295B}). 
At so low temperatures the admixture of helium should produce substantially different CIA effect due to H$_2$--He collisions and therefore would be different from pure hydrogen atmosphere
spectrum in the IR, as indicated in Figure 5 of \citet{2008AJ....136...76H}. In this study we present the full spectral energy distribution
of the companion, and by obtaining good match with the pure-H atmosphere model we can be confident
that the object is indeed a H-atmosphere WD. This allows for proper derivation of the atmosphere
parameters ($T_{\rm eff}$, $g$) and the age of the WD. 
Measurements of near- and -mid IR fluxes are essential for a definitive assignment of the atmosphere composition of other cool WD pulsar companions.
The WD companion atmosphere of PSR J0751+1807 was classified as helium-rich by \citet{2006A&A...450..295B}, which is at odds with the evolutionary models for such stars.
In that particular work such a conclusion was reached by performing analysis with the WD atmosphere models of
\citet{1995ApJ...443..764B}. However, these models tend to bias the prediction of the atmospheric
content of a cool white dwarfs from hydrogen-rich
to helium-rich , as they do not fully account for the red
wing of the Ly-$\alpha$ absorption and the free-free opacity of
dense helium (see discussion in \citet{2006ApJ...651L.137K}. Our conclusion for the atmospheric composition of WD companion of J0437$-$4715,
obtained by a good fit to the entire WD spectral energy distribution by the pure-H atmosphere model,
is in line with other works, which postulate a pure hydrogen outermost layer
\citep{1998MNRAS.294..569H,2001AN....322..405S}. 


Using this (now fixed) model for the WD spectrum, we can again fit a PL to the pulsar part of the spectrum. We choose to fit the range $\lambda<3575$\,\AA, where the pulsar dominates. The fit is to be compared with the black-body fits in Table \ref{bb_pl}; we find $\alpha=-2.46(17)$, $N_{\rm PL}=4.61(19)\times10^{-18}$\,erg\,s$^{-1}$cm$^{-2}$\AA$^{-1}$. From Figure \ref{WD_PL}, it would appear that the fit is good, except for the portion of the spectrum in the NUV, where the WD atmosphere and the PL are of similar strengths. Also, the value of the power-law index is somewhat puzzling, since it is no longer thermal-like, nor is it typical for a pulsar non-thermal spectrum. Indeed, the residuals in the FUV appear to show a systematic slope. This suggests that the UV spectrum may actually consist of two components, for instance a thermal Rayleigh-Jeans tail (PL with $\alpha=-4$) and a non-thermal PL more typical of old MSPs (PL with $-1\geq\alpha\geq-2$, i.e., photon index $1\leq\Gamma\leq2$) - see Section 4.


\subsubsection{Thermal FUV}

Using HST/STIS, K04 were able to measure the FUV spectrum in 4 bins with low S/N, which rendered the measurement of the spectral slope rather uncertain. 
Our much better quality FUV spectrum from ACS matches those STIS fluxes rather well (Figure \ref{spec}) and also the FUV broad-band photometry. This gives us confidence in the accuracy of the cross-calibration between the instruments.

Under the assumption that the FUV PL-like emission is indeed a Rayleigh-Jeans law ($F_\lambda\propto\lambda^{-4}$ or $F_\nu\propto\nu^2$), we can constrain the temperature and radius of the emitting body for the known distance, $d=156.3$\,pc. We calculate $TR^2 = 1.9\times10^7$\,K\,km$^2$.  In the case here, however, a black-body law starts to deviate from a Rayleigh-Jeans law in the FUV for temperatures $T<5\times10^5$\,K. Note that $R\equiv R_\infty$ and $T\equiv T_\infty$, i.e., the temperature and radius as measured by a distant observer.

\begin{figure}
\includegraphics[width=\hsize]{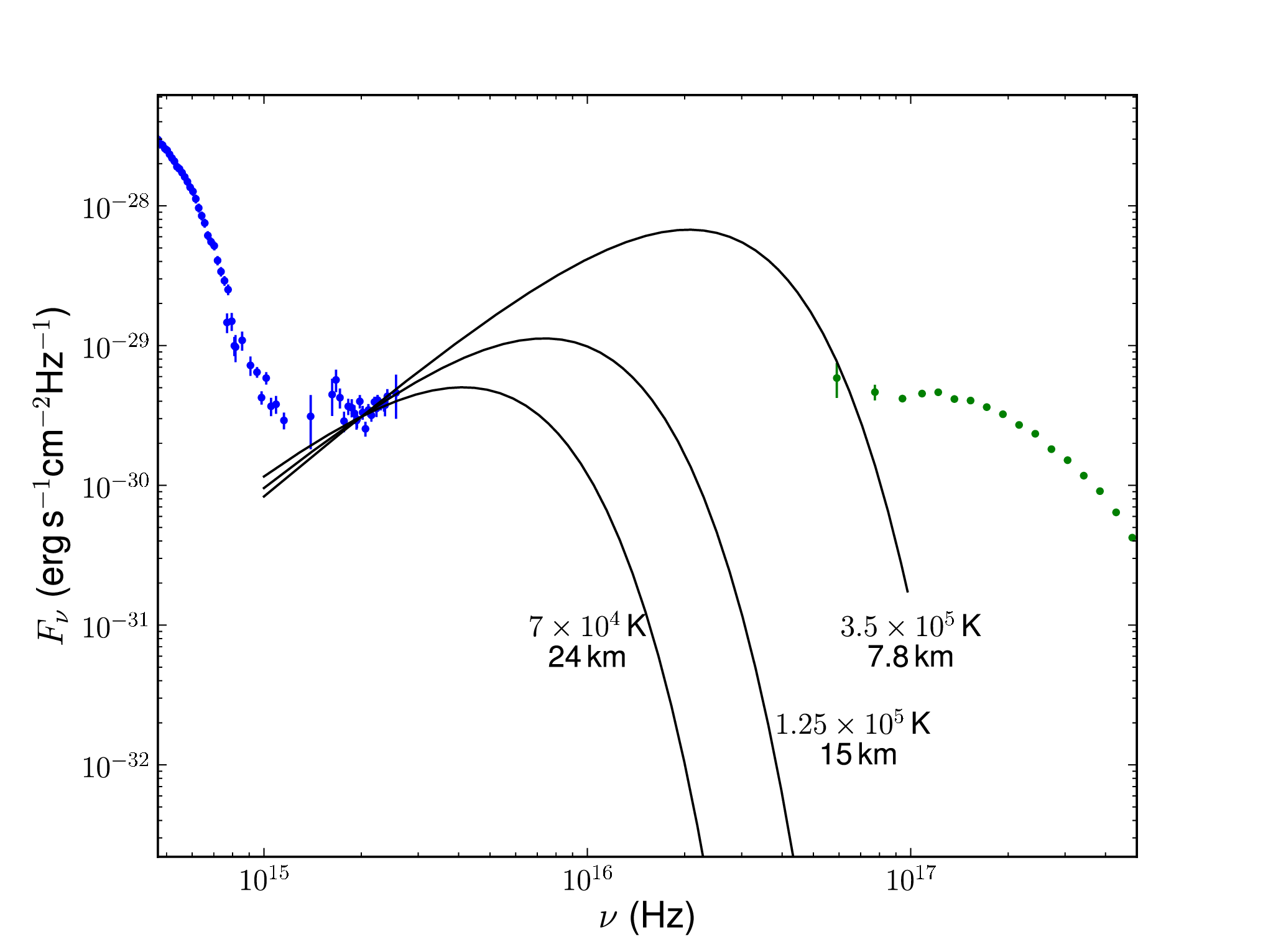}
\caption{Thermal fits to the de-reddened FUV spectrum (points and error bars, left), showing the coolest temperature allowed by the data ($T=7\times10^4$\,K) but with a radius too large for a NS; the coolest temperature compatible with a NS-sized emitter of $R=15$\,km, $T=1.25\times10^5$\,K; and the hottest temperature allowed by the X-ray data (extinction-corrected points and error bars, right), $T=3.5\times10^6$\,K with an emitting radius of 7.8\,km. These values correspond to UV reddening $E(B-V)=0.03$ and X-ray extincting column $N_H=7\times10^{19}$\,cm$^{-2}$.}\label{BBs}
\end{figure}

We show possible thermal fits to the FUV spectrum in Figure \ref{BBs}. We include the lowest-energy points from the X-ray spectra in Section \ref{rawX} -- these constrain the high-temperature limit of the possible thermal emission to $T\leq3.5\times10^6$\,K. The lower limit on the temperature is set by requiring that the emitting area be reasonable for a NS. Whereas temperatures as low as $7\times10^4$\,K can still roughly fit the data, the required radius, $24$\,km, is larger than the largest plausible value of 15\,km \citep{2010NewAR..54..101L}. For $R=15$\,km, the corresponding surface temperature is $T=1.25\times10^5$\,K, and  the likely NS size of $R=13$\,km yields a surface temperature of $1.5\times10^5$\,K.

All of these estimates are affected by the choice of reddening in the FUV and extinction in X-rays (for the upper limit). All the temperatures quoted are for $E(B-V)=0.03$, and should be corrected upwards or downwards by about 15\% for $E(B-V)=0.05$ and $E(B-V)=0.01$, respectively. Similar to the WDA+PL model in the previous section, any WDA+thermal model leaves noticeable residuals in the NUV range (see Figure \ref{WD}).

\subsubsection{Emission Line?}
\begin{figure*}
\begin{center}
\includegraphics[width=0.7\hsize]{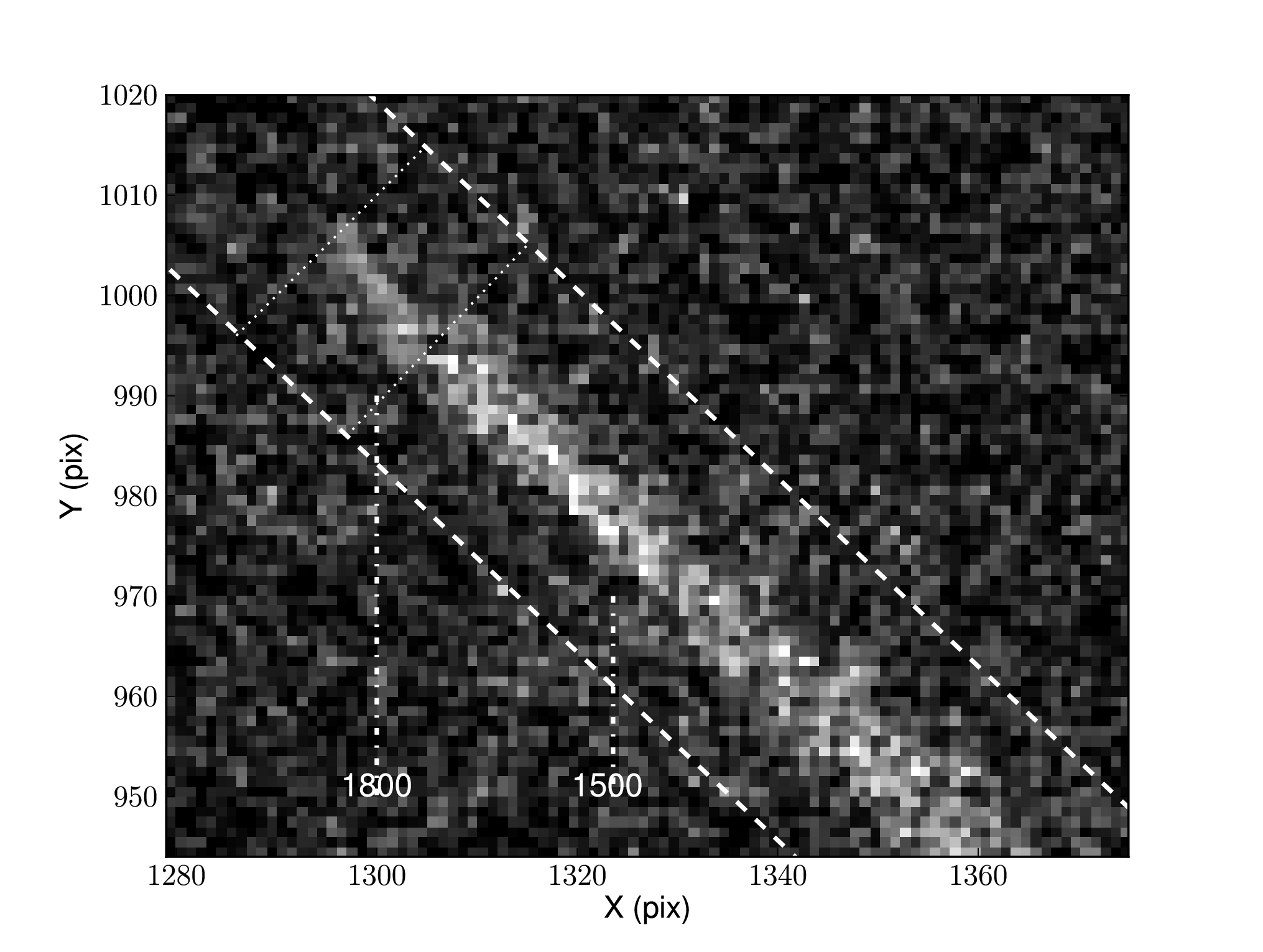}
\caption{Image of the PR130L spectrum, in the region of the possible line at 1800\,\AA. The approximate extent of the feature is marked by white dotted lines. }\label{line}
\end{center} 
\end{figure*}

In Figure \ref{analyt} enhanced flux seen in the residuals around 1800\,\AA, attributable to three bins of the FUV spectrum. Figure \ref{line} shows a zoom of the combined spectrum image, with enhanced contrast, and the vicinity of the line marked. No feature is seen in the raw data by eye; the spectral response is, however, rapidly falling to longer wavelengths in this region. If the point at 1795\,\AA\ were omitted, there would be no question of  a line; as it is, we cannot rule out for certain the existence of an emission line here. We do not, however, attempt to fit a line to the three points, just note that their presence may slightly affect our accuracy in measuring the slope of the spectrum 1000--1700\,\AA.

No evidence is found of a spectral line at 1372\,\AA, suspected by K04. Although there does happen to be an upward fluctuation in the gross counts spectrum at approximately the right wavelength, it is consistent with a fluctuation in the background, and smaller in the final flux spectrum than either other fluctuations or the uncertainty expected for Poissonian statistics. As K04 claimed only a marginal detection, we assume it was due to a statistical fluctuation only. 

\subsection{X-ray band}\label{rawX}
We fitted the set of X-ray observations simultaneously using the X-ray analysis package {\sl Sherpa}. The eight separate spectra were loaded, together with their corresponding response and background data. Each spectrum was grouped into bins with a signal-to-noise ratio of least 8 for {\sl XMM} and 5 for {\sl Chandra}. All fits were performed on the whole data-set simultaneously, with energy bounds 0.1--10\,keV for {\sl XMM} MOS, 0.6--10\,keV for {\sl XMM-Newton} PN and 0.3--8\,keV for {\sl Chandra} ACIS data.

The results of fits with a variety of continuum models is shown in Table \ref{Xfit}.
Among others, we used the neutron star atmosphere model implemented in Xspec as ``NSAgrav'' for non-magnetised atmospheres \citep{1996A&A...315..141Z}. Keeping the mass fixed at 1.76\,M$_\odot$ \citep{2008ApJ...679..675V}, and the radius at the default 10\,km, we find a  marginally better fit than with a single BB. The hot component best-fit normalization corresponds to a large distance ($d>5$\,kpc) if the emission were coming from the entire visible surface. Since the true distance is $d=156$\,pc, this suggests that the emission is coming from a region with effective radius less than 3\% of the neutron star radius.

\begin{figure*}
\begin{center}
\includegraphics[width=0.8\hsize]{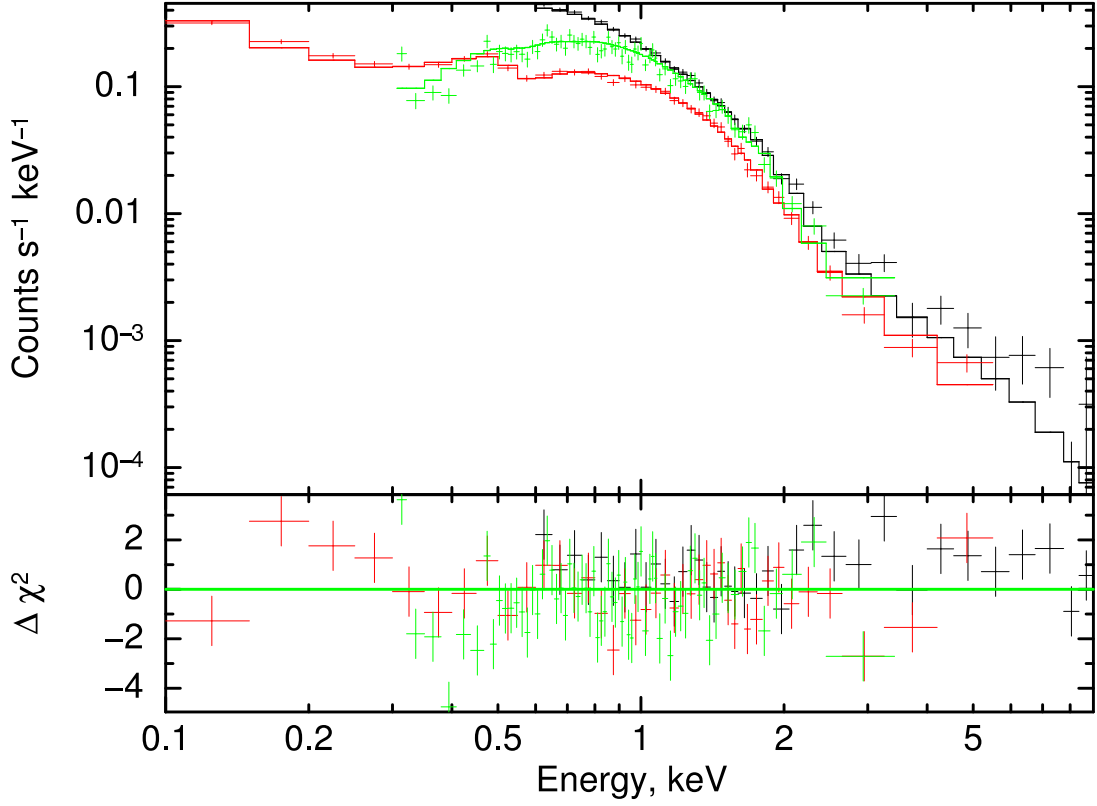}
\caption{Count-rate spectrum of \psr, with NSAgrav+PL model and $N_H=7.4\times10^{19}$\,cm$^{-2}$ extinction, which had the lowest reduced chi-squared value of our fits. The black points are PN data, green is ACIS and red is MOS1.}\label{counts_spec}
\end{center} 
\end{figure*}

The upper limit established by \citet{1998A&A...329..583Z} on the column density to \psr\ is 1.2$\times10^{20}$\,cm$^{-2}$, based on the X-ray spectrum of the nearby AGN. This would correspond to $E(B-V)\approx0.02$ optical reddening \citep{1995A&A...293..889P} (note that this relation shows considerable scatter, especially for low extinction). From fitting the softer {\sl ROSAT}/PSPC spectra with black-body, power-law or neutron star atmosphere models yielded column densities in the range (1--9)$\times10^{19}$\,cm$^{-2}$. We performed two sets of fits, one with the  fixed  hydrogen column density of $N_H=7\times10^{19}$\,cm$^{-2}$, and a second set with the extinction free. We used the cross-sections of \citet{1983ApJ...270..119M}, as implemented in the model {\tt wabs}. The effect of extinction for $E>0.5$\,keV is very small. The fit parameters we obtain are generally similar between the fixed and free extinction parameter. In Figure \ref{counts_spec} we show one of our fits, with the NSAgrav+PL model and fitted extinction. 

\begin{figure*}
\begin{center}
\includegraphics[width=0.7\hsize]{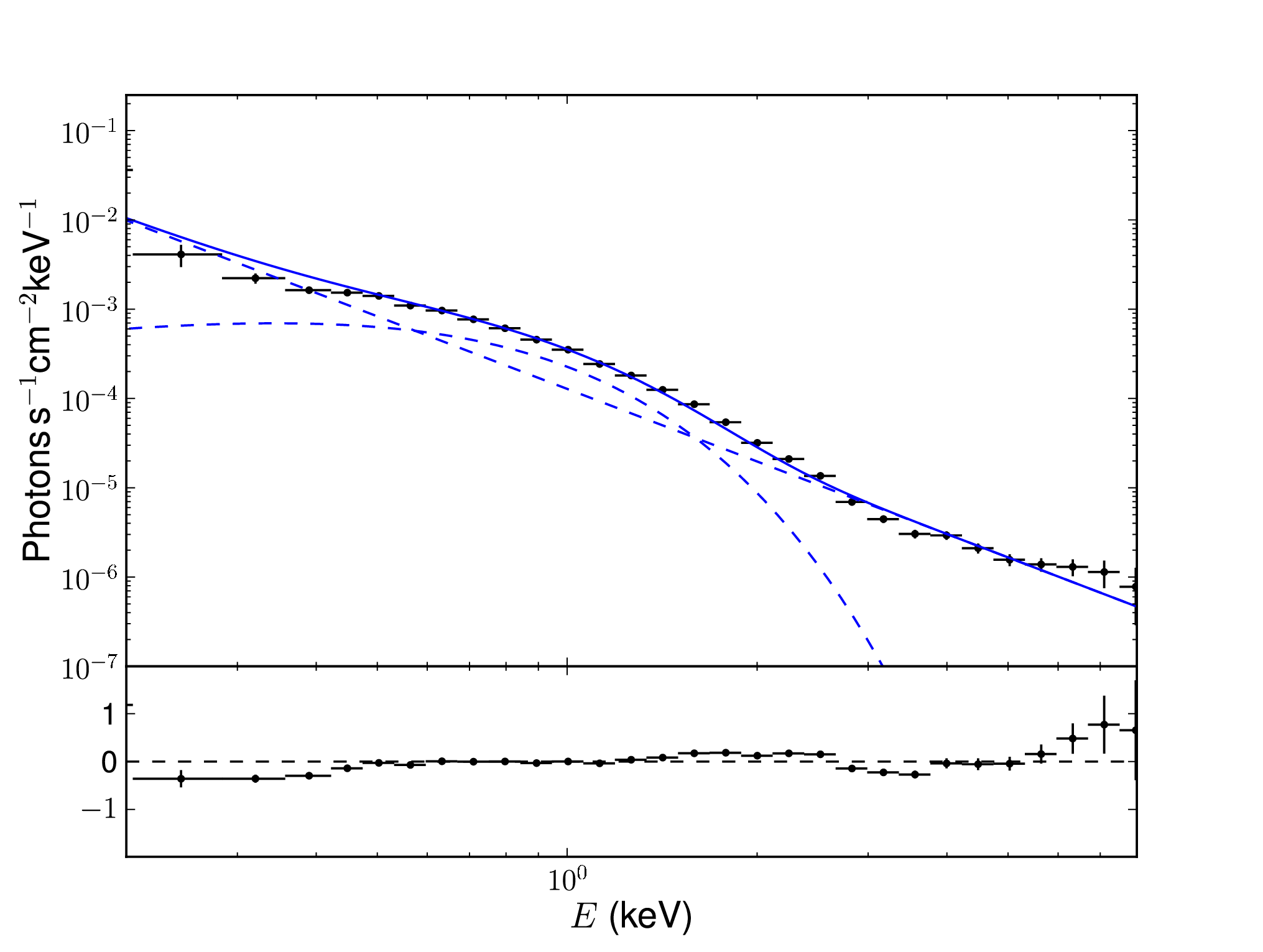}
\caption{The X-ray BB+PL model, compared to the combined fluxed spectrum (i.e., each flux point is the weighted average of the fluxes from the three {\sl XMM-Newton} and five {\sl Chandra} spectra in that energy bin). The model was obtained by standard `forward fitting' of the count rate spectra. Fractional residuals are shown in the bottom panel (i.e., (f-model)/model). }\label{Xfitgraph}
\end{center} 
\end{figure*}

In Figure \ref{Xfitgraph} we show the average fluxed spectrum (see Section 4), plotted with the best-fit BB+PL model. We can see that generally the PL component's photon index is not only determined by the high-energy tail, but by a combination of this tail and the low-energy, extinction-affected data. Since the PL cannot rise so steeply into the EUV/FUV, we suspect that the X-ray spectrum  is better modeled by multiple thermal components and a flatter PL. Furthermore, although the fit is formally acceptable, the majority of the $\chi^2$ is contributed by the high signal-to-noise 0.5--2\,keV range, leaving some structure in the residuals. 

The variety of models tested in Table \ref{Xfit}, all produce statistically acceptable fits, except BB+BB. We see thermal emission dominating the high signal-to-noise bulk of the spectrum with possibly further cool thermal temperatures at the low energies and PL at high energies. 
If we perform the fitting only for the higher energies, we find increasingly small (but poorly constrained) values of the photon index; e.g., for BB+PL fitted above 1\,keV, $\Gamma=2.2(3)$, and for PL only fitted to energies above 3, 4, and 5\,keV, we find $\Gamma=1.6(4)$, 2.0(6) and 1.6(9), respectively. Since the two-temperature atmosphere models also yield similar values of the photon index (Table \ref{Xfit}),  it is more likely a more reallistic description of the high-energy spectrum.

\begin{deluxetable*}{ccccccccc}
\tablecaption{X-ray Spectral Fits\label{Xfit}}
\tablewidth{0pt}
\tablehead{
\colhead{Model} & \colhead{$N_H$}& \colhead{$kT_1$} & \colhead{$R_1$} & \colhead{$kT_2$} & \colhead{$R_2$} & \colhead{$\Gamma$} & \colhead{$A\,^1$}
 & \colhead{$\chi^2/dof$}  \\
 &\colhead{(10$^{19}$\,cm$^{-2}$)}  & \colhead{(eV)} & \colhead{(km)} & \colhead{(eV)} & \colhead{(km) }
}
\startdata 
BB+PL & 7 &216(2) & 0.073(3) &\ldots &\ldots & 2.699(19) & 1.28(3) & 646/629\\
BB+BB & 7 &286(5) & 0.045(5) & 108(3) & 0.35(4) & \ldots&\ldots & 1084/629\\
BB+BB+PL&7&258(12)& 0.044(7) & 141(13)& 0.112(15)&2.69(4)&1.05(4)& 598/627\\
NSAgrav+PL& 7 &138(2)&0.294(12)& \ldots& \ldots& 2.69(3) & 1.05(3) &  601/629\\
NSAgrav+NSAgrav+PL&7&140(20)&0.4(2)&30(11)&7(2)&1.9(6)&0.31(2)$_{-0.2}^{+0.6}$&  599/627\\
\hline
BB+PL & 13.9(13) & 223(3)& 0.065(2) & \ldots& \ldots& 2.86(3) & 1.46(4) & 613/628 \\
BB+BB+PL & 8$_{-2}^{+5}$& 240$_{-10}^{+60}$ & 0.05$_{-0.04}^{+0.002}$ &125$_{-4}^{+60}$&0.15$_{-0.1}^{+0.02}$&2.69$_{-0.08}^{+0.2}$&1.1(2)& 599/626\\
NSAgrav+PL & 7(3)& 140(14) & 0.36(2) &\ldots &\ldots & 2.68(5) & 1.05(9) & 599/628\\
NSAgrav+NSAgrav+PL & 2.5$_{-1.5}^{+2.5}$ & 158(13) & 0.24(5) & 48(9) & 2.2(8) & 1.6(4) & 0.23(19) & 592/626
\enddata
\tablecomments{PL: power-law, BB: black-body, NSAgrav: neutron star hydrogen atmosphere (non-magnetized). Numbers in parentheses show the one-sigma confidence interval in the last digit(s). Models in the upper section have fixed column density, $N_H$, and in the lower section it is allowed to vary. For NSAgrav, $T=T_{\rm eff}$, the model effective temperature (uncorrected for red-shift) and the radius $R$ of the effective emitting area.\\
$^1$ PL normalization $A$ in units of $10^{-4}$\,ph\,s$^{-1}$cm$^{-2}$keV$^{-1}$ at 1\,keV. }
\end{deluxetable*}

\subsection{Gamma-rays}\label{gam}

We modeled the spectrum in the {\sl Fermi} band by direct maximum-likelihood analysis of the calibrated photons falling near the position of \psr\ on the sky.

In order to derive spectral models and fluxes for J0437, we followed the standard procedure for {\sl Fermi} LAT data\footnote{ See http://fermi.gsfc.nasa.gov/ssc/data/analysis/scitools/}.
 In particular we modeled out emission from 
 1FGL  sources (see Abdo et al. 2010a) within a $r=10^{\circ}$ circle centered on J0437.
We derived the spectral fit to the J0437 spectrum in 0.1-4 GeV range\footnote{There are virtually no photons detected from J0437 above 4 GeV.} by  maximizing  the unbinned likelihood with the {\tt gtlike} 
  procedure.  Standard background models were used to account for both Galactic and extragalactic diffuse emission as well as instrumental background. We have fitted both PL and cutoff PL [$E^{-\Gamma}\exp(-E/E_{\rm cut})$] models. Due to the limited statistics available in the data, the  cutoff energy was not well constrained by the fitting, with a best-fit value of $E_{\rm cut}$=1.3\,GeV, but with a broad acceptable range. We obtain best-fit $\Gamma=2.6\pm0.2$  and $\simeq2.2$, for the PL fit and the cut-off PL fit, respectively. The observed flux in 0.1-100 GeV is $2.7(8)\times10^{-11}$\,erg\,s$^{-1}$cm$^{-2}$ (compared to the 1FGL value $2.4(7)\times10^{-11}$\,erg\,s$^{-1}$cm$^{-2}$). With the statistics available it is not possible  to firmly discriminate between the   PL and cut-off PL models, but better S/N sppectra of other MSPs suggest that cut-offs above 1\,GeV are ubiquitous in MSPs \citep{2010ApJS..187..460A}.
  
The slight difference between the fit parameters here and those listed in \citet{2010ApJS..187..460A} are likely due to the larger amount of data which we analyzed.
  
In order to produce fluxes and uncertainties for the multi-wavelength fitting below, we spit the available energy range into five bins, and repeated the  {\tt gtlike} procedure for each bin, assuming  cut-off PL. 

\section{Multi-wavelength analysis}\label{MWan}
We wish to consider whether the ensemble of data across many orders of magnitude in photon energy can  be fit by simple models.

We converted each spectrum into spectral energy flux, $F_\nu$. This involves dividing the number of counts in a bin with the effective area at that particular frequency and multiplying by the bin energy. In the optical/UV regime, the effective area  is simply the sensitivity function (also called zero point for photometry), but at higher energies, one normally also considers {\sl spectral redistribution}: the number of counts in a given energy bin is determined by the incident spectral flux not only at that energy but also at other energies. Whereas in the optical this is typically only from the neighboring bins (and sets the sample spacing in spectral space), for X- and $\gamma$-rays, a photon of given energy may generate signal at a very different energy. We define the flux $F$ in bin $h$ as (see e.g., \citealt{2005Ap&SS.300..159N}):
\begin{eqnarray}
F_{\rm fluxed}(h) = \frac{\left[ C(h)-B(h) \right]/\Delta t}{ \int R(h,E)A(E)dE}
\end{eqnarray}
where $C(h)$ and $B(h)$ are the gross counts and background in bin $h$, $R(h,E)$ is the redistribution matrix i.e., the normalized probability of a photon of energy $E$ being detected in bin $h$, and $A(E)$ is the effective area for energy $E$.

In order to combine the X-ray spectra, we sample the flux from each data-set in a given energy bin, and take the mean weighted by the uncertainties.  We note that although redistribution undoubtedly affects the appearance of narrow features, for the wide spectral range and broad bins considered here, it should make  little difference.


\begin{figure*}
\begin{center}
\includegraphics[width=0.8\hsize]{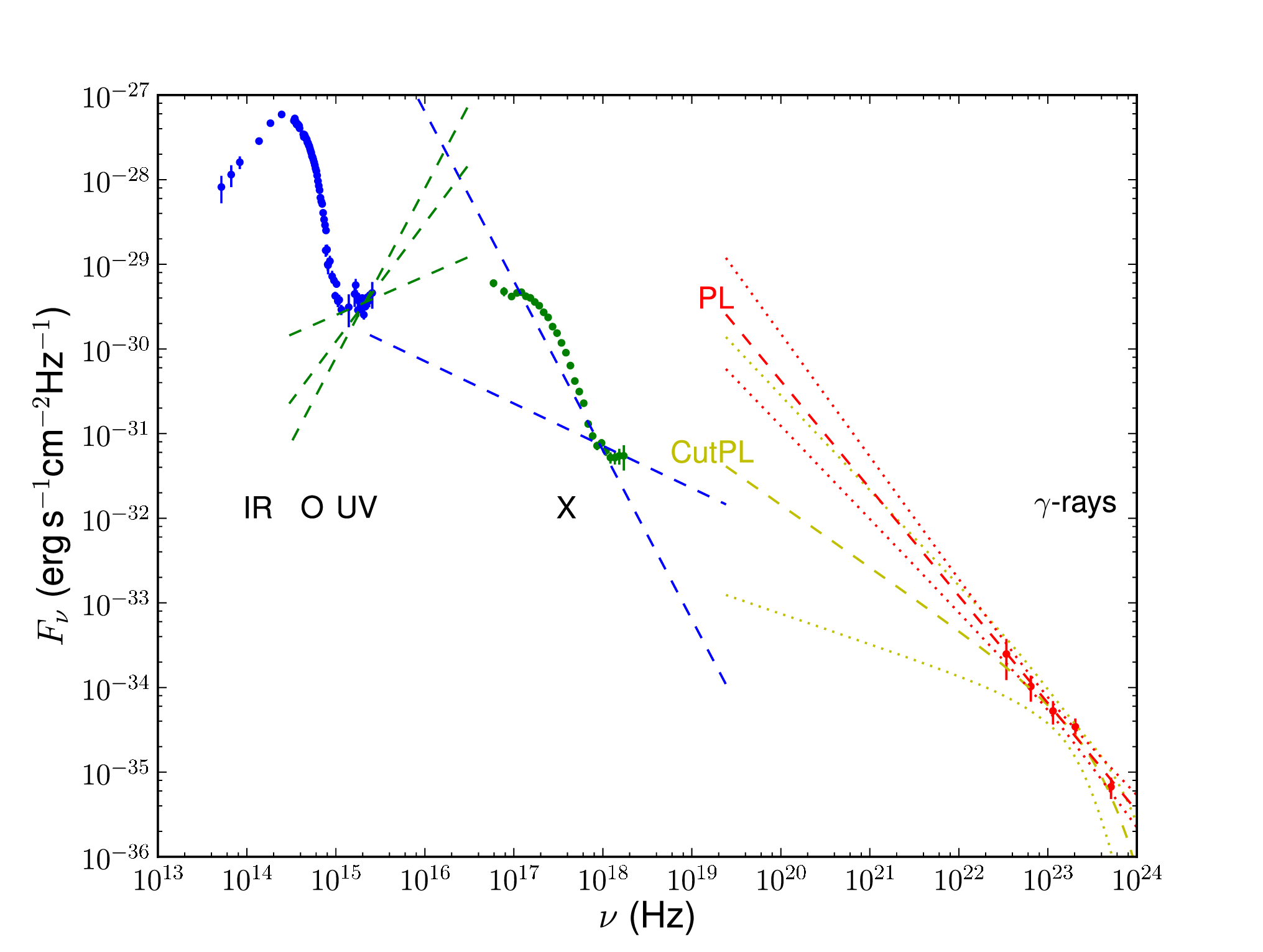}
\caption{Multi-wavelength spectral energy distribution of \psr, showing the raw data (points with error bars).   Over-plotter are curves showing fits from the different wavebands. For the $\gamma$-ray band, we show both a cut-off PL and pure PL fits, with dashed lines for the best fits, and the 1-$\sigma$ confidence bounds shown with dotted lines. For the X-rays and optical-UV , we show the range of PL components from the various  models, where the formal uncertainty in each PL slope is small compared to the differences between models ($\alpha=-2.5,-3.4,-4$ for the optical and $\Gamma=1.5,3$ for the X-ray). The thermal (pulsar and WD) components are not shown for simplicity.}\label{MW2}\label{MW1}
\end{center}
\end{figure*}

In Figure \ref{MW2} we show the  multi-wavelength spectrum of \psr\ and the fits that were made for the spectral bands separately: the IR-UV, X-rays and $\gamma$-rays. The fits are combinations of PLs and thermal contributions, but we show only the PL components, extrapolated over many orders of magnitude, because a thermal component fitted to the X-rays, for example, does not contribute in any other spectral window. For the $\gamma$-ray fit to the {\sl Fermi} data, there is significant uncertainty in the best-fit parameters. In order to calculate the range of spectra consistent with the models, we perform 10000 simulations of spectra, taking parameters from Gaussian distributions with central value and spread given by the fitting; dashed lines in Figure \ref{MW2} show the best-fit models, and dotted lines represent the one-sigma confidence bounds. For extrapolating the X-ray and optical spectra, the PL photon indices are very well defined for a given fit, but fits with different models give different, inconsistent values. We plot the best-fit  PL components of the various models, $\Gamma=3, 1.5$ for X-rays and $\alpha=-4, -3.4, -2.5$ for the UV. Here the value $\alpha=-4$ corresponds to a Rayleigh-Jeans law (note that  $\alpha=\Gamma-3$).

We see from Figure \ref{MW2} that most of the extrapolations are not  consistent between the spectral bands, with the  exception of the cut-off PL in $\gamma$-rays and the flattest PL in X-rays. Note that a steep PL fit ($\Gamma=3$) in X-rays would need to break sharply above FUV frequencies in order to connect the X-ray and FUV spectra. The PL-only fit to the {\sl Fermi} data also appears unlikely, as it would need to turn over strongly as it approaches the X-ray band. Similarly, the PL fits to the FUV spectrum must turn over before the softest energies of the X-ray band, but this would be expected for thermal emission. 

\begin{figure*}
\begin{center}
\includegraphics[width=0.8\hsize]{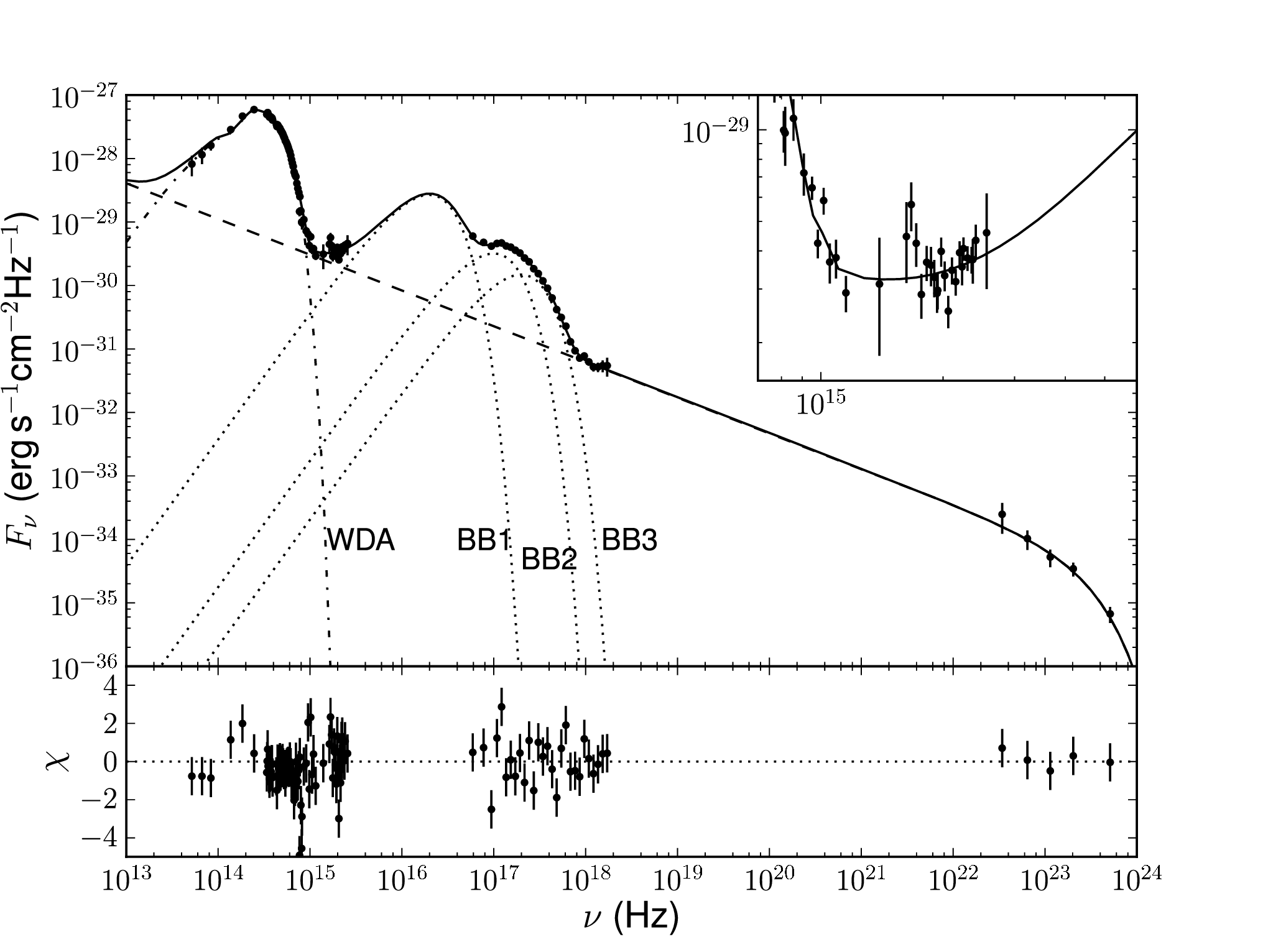}
\caption{Multi-wavelength fit to the spectral energy distribution of \psr, with WD atmosphere model, black-bodies (labeled BB) and a cut-off power-law ($\Gamma=1.563(11)$, $\nu_{\rm cut}=2.7(5)\times10^{23}$\,Hz). The inset shows a zoom of the UV part of the spectrum.}\label{MW3}
\end{center}
\end{figure*}

\begin{figure*}
\begin{center}
\includegraphics[width=0.8\hsize]{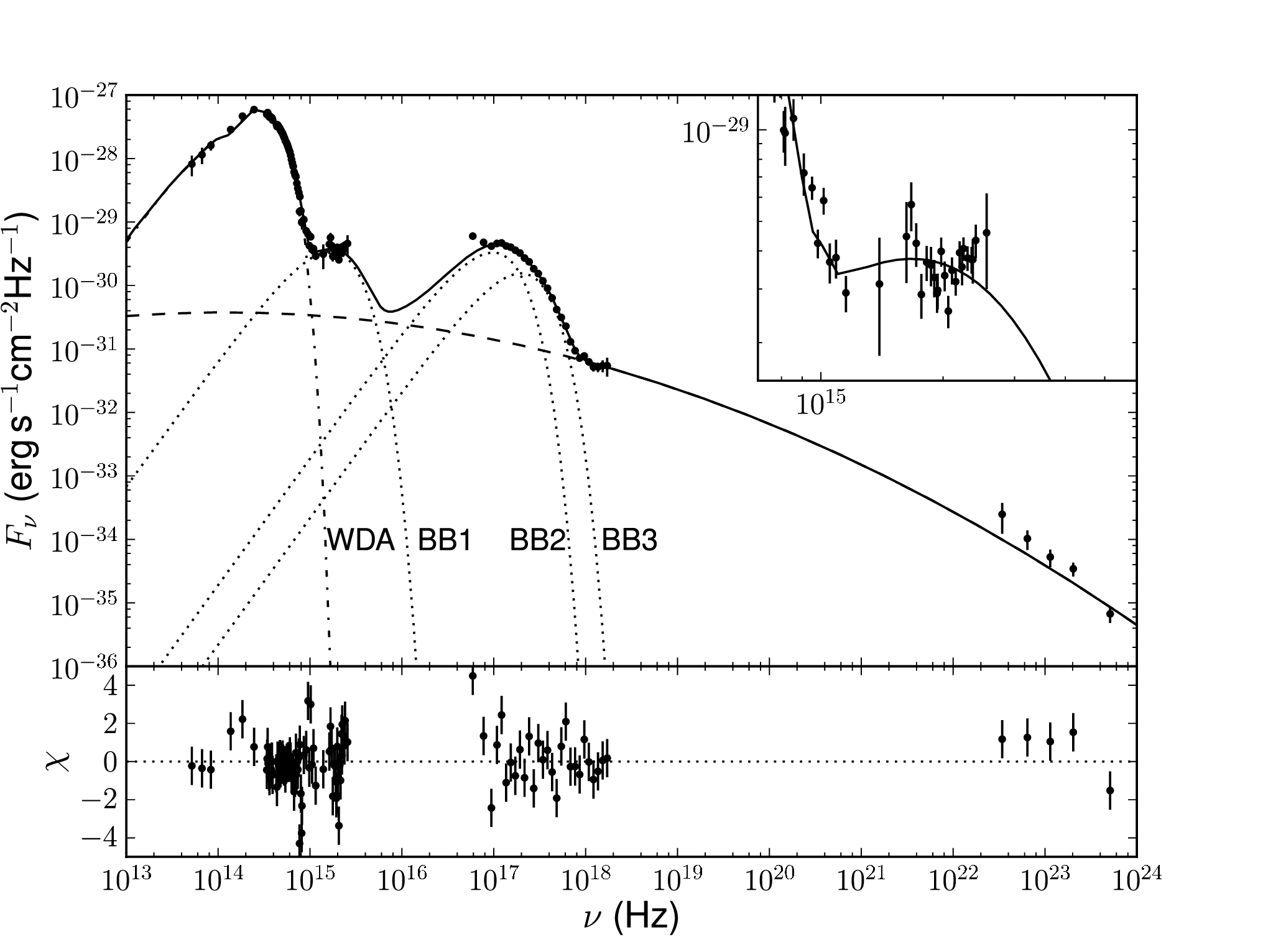}
\caption{Multi-wavelength fit to the spectral energy distribution of \psr, with WD atmosphere model, black-bodies (labeled BB) and a variable-index PL ($a=0.89$, $b=0.05$ in Equation 2). The inset shows a zoom of the UV part of the spectrum.}\label{MW4}
\end{center}
\end{figure*}

It turns out to be possible to fit simple models to the whole set of multi-wavelength spectral data. For example, in Figure \ref{MW3}, we have fitted a single cut-off PL across the whole range, with a number of thermal components (black-bodies for the UV-X-rays, and the same WD atmosphere as before in the IR/optical). 
Since in this case, the WD part is fixed, we only fit for frequencies $\nu>9\times10^{14}$\,Hz ($\lambda<3300$\,\AA). The optical reddening is fixed $E(B-V)=0.03$ and the X-ray extinction $N_H=7\times10^{19}$\,cm$^{-2}$. We find that the entire pulsar  spectrum (excluding WD) is reasonably well fit ($\chi^2_\nu \approx 1.2$) by a single cut-off power-law and three black-bodies. The fit parameters are given in Table \ref{MWfit} (see also  Figure \ref{MW3}). The uncertainty on the coolest temperature is small because the that component contributes in both the FUV and soft X-rays, whereas in principle additional thermal components could be included in the fit; we note that the radius for this component is somewhat too small for a neutron star.  With the gap in the spectrum between the FUV and soft X-rays, we are thus not able to constrain the surface temperature better than in \S 3.1.3, $1.25\times10^5\leq T\leq3.5\times10^5$\,K. Further data at the softest X-ray energies ($<0.2$\,keV) and/or an independent measurement of the X-ray extinction would be very useful. The latter could possibly be obtained from measures of photo-electric edges in high-resolution grating spectra (e.g., \citealt{2006ApJ...650.1082D}).

We find that thermal emission from the bulk of the neutron star surface, with a temperature of 1.5$\times10^5$\,K is supported by the data. Furthermore, we have derived a rough temperature distribution on the surface of the neutron star, largely cool, but with a much hotter polar cap (or caps). In this model, it seems plausible that the cut-off PL, which can describe the $\gamma$-ray emission adequately, is a continuation of the same PL detected in the X-rays, and which indeed also makes a contribution in the NUV range. 

An alternative to the cut-off PL above could be a variable-index PL; e.g., we know from multi-wavelength observations of young pulsars that the spectrum often softens with increasing energy \citep{2007Ap&SS.308..287K}, which can be modeled as 
\begin{equation}
F_\nu = N\nu_{23} ^{-a-b\log\nu_{23}} ,
\end{equation}
where $\nu_{23} = (\nu/10^{23}$\,Hz) . This model has the advantage of producing a  near flat  $F_\nu$ in the optical regime, as is seen for non-recycled pulsars with higher $\dot{E}$  but steeper at higher energies\footnote{An exponential cut-off is still expected in the GeV range, but with the steeper model spectrum, the fit is no longer sensitive to it for our data.}.
In our case, we can either require the PL component to contribute only in $\gamma$-rays and X-rays by fixing $b=0.05$ (Figure \ref{MW4}), in which case $a=0.89(2)$ (see Table \ref{MWfit}). If we allow $b$ to vary, the best-fit curvature is $b=0.013(3)$ and the remainder of the parameters tend toward those for the cut-off PL fit.
Both of the variable index fits give statistically worse $\chi^2$ than the cut-off PL above.

\begin{deluxetable*}{lcccccc}
\tablecaption{Multiwavelength fits\label{MWfit}}
\tablewidth{0pt}
\tablehead{
\colhead{Fit} & \colhead{$T_{\rm BB}$ ($10^6$\,K)} & \colhead{$R_{\rm BB}$ (km)} & \colhead{$\Gamma$}  &  \colhead{$F_\nu(10^{23}\,{\rm Hz})$} & \colhead{Other parameters} & \colhead{$\chi^2/dof$}
}
\startdata 
CutPL+3BB & 3.37(9),1.70(5),0.34(2) & 0.039(4),0.160(13),5.3(8) & 1.562(13) &0.99(18) &$\nu_{\rm cut}=2.7(6)\times10^{23}$\,Hz & 79/54\\
VarPL+3BB & 3.36(7),1.66(4),0.0285(15) & 0.040(5),0.168(10),76(4) & 1.893(16)& 3.9(6) &$b=0.05$ (fixed) &130/55\\
VarPL+3BB & 3.35(7),1.69(4),0.337(12) & 0.039(6),0.160(11),5.2(8) & 1.73(3) & 3.0(4) & $b=0.013(2)$& 93/54
\enddata
\tablecomments{Black-body and power-law fit parameters for multi-band models: CutPL (cut-off PL) and VarPL (PL with a variable index, see text). Numbers in parentheses indicate 1-$\sigma$ uncertainties in the final digit(s). For the VarPL fits, $\Gamma=a+1$.}
\end{deluxetable*}


\section{Discussion}

With the amount of MW data gathered and analyzed,  J0437 has  the most comprehensive MW spectrum of an MSP to date.  Also, we have measured the mid-IR-NUV spectrum of the WD companion, where, thanks to the precisely determined distance, mass and full spectral energy distribution, the WD is one of the best-characterized WDs known. 

%
  
\subsection{WD companion}
The modeling in \S \ref{WDat} demonstrates the high accuracy of the latest WD atmosphere models. We find that the WD has a radius  $R=(1.87\pm0.20)\times10^4$\,km and temperature $T=3950\pm150$\,K and a  hydrogen surface, for a WD of mass $M=0.254$\,M$_\odot$. It is one of the coolest WDs observed to date (e.g., \citealt{2004ApJ...612L.129G,2008AJ....136...76H,2010ApJ...715L..21K}).

The WD radius is about 40\% larger than the theoretical zero-temperature, zero-rotation radius expected for a pure He WD \citep{1961ApJ...134..683H,1998A&A...339..123D}. This can be seen as a result of the non-zero temperature hydrogen envelope and the unknown rotation rate. Very high internal magnetization ($B\sim10^{12}$\,G) of the WD (where the surface field is only a small fraction of the interior field strength) could also result in an additional radius increase of up to 10\% \citep{2000ApJ...530..949S}.
  
\citet{1998MNRAS.294..569H} considered the \psr\ system in detail, in order to derive limits on the age and initial pulsar spin period (since the cessation of accretion). The temperature, surface composition and mass of the WD companion to \psr\ can be used to derive an estimate on its age. Using Figure 2 of \citet{1998MNRAS.294..569H}, we estimate $\tau_{\rm WD} = 6.0\pm0.5$\,Gyr. 
For a 0.242\,$M_\odot$ mass WD, \citet{2001MNRAS.325..607S} also predicted a surface temperature and colors consistent withour findings after 6\,Gyr.

It is interesting to compare the inferred WD age with the pulsar age.
A correction must be applied to the apparent characteristic age of the pulsar, caused by the transverse Doppler effect due to the pulsar's transverse velocity (i.e., proper motion, \citealt{1970SvA....13..562S}). For the particular case of \psr, being close to us and having a large proper motion, this correction is significant. Specifically, the Shklovskii correction to the characteristic rate for proper motion, $\mu$, and distance, $D$, is given by $(\dot
P/P)_i = (\dot P/P)-\mu^2 D/c$ 
where the subscript $i$ denotes the intrinsic value.
Using the precise VLBI measurements of $\mu$ and $D$ by \citet{2008ApJ...685L..67D},  we find the corrected spin-down characteristic age is $\tau_{\rm PSR} = \frac{P}{\dot{P}_i(n-1)} \left[ 1 - \left( \frac{P_0}{P} \right)^{n-1} \right] = 6.7\pm0.2$\,Gyr, assuming breaking index $n=3$ and spin period $P_0\ll P=5.8$\,ms.

The derived ages are rather similar with one-another, surprising given the number of assumptions involved. This implies $P_0$ must be substantially faster (at least a factor of 1.4 for $n=3$) than the current period of 5.8\,ms (cf., 1.4\,ms for the fastest known pulsar; \citealt{2006Sci...311.1901H}).

  \subsection{Pulsar thermal emission}
  
We have constrained the temperature of the bulk of the pulsar surface to $1.25\times10^5\leq T \leq 3.5\times10^5$\,K for reasonable radii, or 1.5$\times10^5$\,K for the preferred radius of 13\,km, with about a 15\% uncertainty dominated by the uncertainty in the extinction. It may be possible to further constrain this temperature by filling in the spectral gap from FUV to soft X-rays, is adequate instrumentation is available. 
  
It is obvious from  our MW fits that the thermal component cannot be described well by a single BB or NS atmosphere model, supporting the conclusions of \citet{1998A&A...329..583Z} and \citet{2002ApJ...569..894Z}. The large width of the thermal peak argues strongly in a favor of  multi-temperature thermal spectrum, which  would be the natural consequence of a non-uniformly  heated NS surface.  If modeled as a sum of BBs, the observed spectrum requires a high-temperature ($\simeq 4$\,MK) component   that appears to be emitted from a relatively small area and it  also requires a cooler component that must be radiated  from  most of the NS surface. Hence, one could suggest that the hot component 
 comes from heated polar caps (see K04 and references therein) while the cool component is due to the residual heat stored in the NS interior.
   However,  the temperature of this cooler component, $\simeq0.15$ MK, is surprisingly  large for a passively cooling NS. This means that either a heating mechanism operates in the NS interior 
   or the heat deposited at the polar caps is efficiently conducted to the rest of the NS surface. 
In modeling the phase-resolved spectrum, \citet{2007ApJ...670..668B}  suggest some axial asymmetry for the hotter components explained by an offset dipole magnetic field.
   
\subsubsection{Heating mechanisms}
   
After an initial neutrino-dominated cooling epoch, passive photon-dominated cooling from the NS surface is expected to rapidly reduce the surface temperature to below $10^5$\,K by 1\,Myr, and $10^3$\,K by 1\,Gyr (see the review by \citealt{2004ARA&A..42..169Y} and \citealt{2006NuPhA.777..497P}). The surface temperature we find is well above this. There may be a number of possible mechanisms for internal re-heating of the NS which can result in higher surface temperatures of very old pulsars. Such mechanisms are discussed in \citet{2006NuPhA.777..497P}, but we briefly summarize several of the relevant possibilities here.
\begin{itemize}
\item{The interior of the NS is thought to contain superfluid, so, as the crust spins down, differential rotation will be created between the layers, and this can release energy by the friction between the `creeping' superfluid vortex lines and the non-superfluid crust to which it is pinned \citep{1999ApJ...521..271L}. Similar processes are thought to be responsible for the well-known glitching behavior of younger pulsars. In late times ($\tau\gg10^6$\,yr), the photon cooling from the NS surface balances the frictional heating, and the thermal luminosity,  independent of the previous NS thermal history, is equal to the spin-down rate multiplied by the excess angular momentum in the core with respect to the crust, $L=J|\dot\Omega|$. \citet{1999ApJ...521..271L} used this to explain the temperatures of three moderately old ($\tau\sim10^7$\,yr) pulsars. }
\item{Ohmic heating  due to currents induced by the NS magnetic field  \citep{2008A&A...486..255A}. This effect is only relevant for high-B NSs. Although the initial field may have been much higher than it is today, all MSPs have weak surface magnetic field, presumably due to processes in the accretion phase. Stronger fields may, in principle, be buried in the NS interior, so we cannot completely exclude ohmic heating.}
\item{The centrifugal force of rapid rotation affects the chemical balance in the NS interior. As the NS spin downs, the force decreases, and the particle populations are forced into non-equilibrium. Chemical reactions restoring this equilibrium will release heat, a process known as rotochemical heating \citep{1995ApJ...442..749R}. For the surface temperature for J0437 reported by K04, \citet{2010AIPC.1265..166P} developed a specific model for heating in this case and concluded that rotochemical heating was the preferred heating mechanism. Our surface temperature range is  slightly higher than that given by K04 moving J0437 slightly further from the rotochemical model cooling curve, but not significantly enough to alter \citet{2010AIPC.1265..166P}'s conclusions. }
\item{With the high mass and density of a NS, it is possible that dark matter particles could be trapped there and either decay or annihilate, releasing energy \citep{2008PhRvD..77d3515B}. The nature of dark matter particles is currently unknown, so the quantitative predictions of such a mechanism remain rather uncertain. Dark matter, even if it liberates both gravitational  and rest-mass energy, probably has far too low a density in the solar neighborhood to contribute significantly to PSR J0437's 1$0^5$\,K surface emission \citep{2010PhRvD..82f3531K}. }
\end{itemize}

Both vortex creep and rotochemical heating mechanisms depend on rotational kinetic energy. According to \citet{2010A&A...522A..16G}, only these two mechanisms will be important at time-scales comparable to PSR J0437's age. Our measurement of the bulk surface temperature, $1.25\times10^5\leq T \leq3.55\times10^5$\,K, is more precise and slightly higher than the temperature found by K04. In Figure 4, top, of \citet{2010A&A...522A..16G}, this pushed the data point from about 1-$\sigma$ from the upper model curve to about 1.5-$\sigma$ from the model curve. Although the allowed range of surface $T$ is in agreement with the more uncertain K04 measurement, we stress that the important result of our analysis is a much increased confidence in the thermal interpretation of the pulsar FUV spectrum (K04 allowed for a non-thermal interpretation as well). 


   
\subsubsection{Polar cap heating}

As well as residual heat in the interior and any heating mechanism discussed above, the surface temperature profile of the pulsar will be affected by the energetics of spin-down induced magnetospheric activity. A large amount of energy is available from the loss of rotational kinetic energy, $\dot{E}_i=2.8\times10^{33}$\,erg\,s$^{-1}$, but the majority of this will be released as a relativistic wind. Some fraction of this power is, however, released as magnetospheric photon emission (which accompany the particle cascades). High-energy magnetospheric particles are responsible for the non-thermal emission from the pulsar, but it is generally believed that some fraction of the particle flux returns to the NS surface along the field lines, heating the polar caps. Indeed, this is believed to be the cause of soft X-ray thermal emission with  $T\ga10^6$\,K and projected areas much smaller than the size of a NS \citep{2007arXiv0710.3517H}. In the case of J0437, it is reasonable to assume that the hotter black-body components fitted above, with a combined luminosity of $L_{\rm hot}\sim10^{30}$\,erg\,s$^{-1}$, are also polar caps heated in a  similar way to other pulsars. The thermal X-ray efficiency of the order $\eta\approx3\times10^{-4}$ is within the expected range \citep{2007MNRAS.376L..67G,2008AIPC..983..171K}.
     
\citet{2007arXiv0710.3517H} reviews the production of particles and radiation in the vicinity of pulsars, and polar cap heating from different  polar cap emission models. Their models for MSPs predict a luminosity similar to that which we find for the `hot' luminosity, $L_{\rm hot}$. Given that these particles will not penetrate beneath the atomic layers \citep{1995JApA...16..375D}, we expect the downward particle flux to have a negligible effect on the temperature observed for the bulk of the NS surface. Heat may, however, propagate along the surface if the conductivity is high enough,  and this is suggested by the broadness of the thermal spectral peak. According to the fits in Table 7, however, the hotpolar cap is small, and the temperatures intermediate between the hottest areas and the bulk of the neutron star surface are emitted from an area still small compared to the total. We therefore infer that the conduction along the surface is probably small.

In short, the surface temperature $1.25\times10^5\leq T\leq3.5\times10^5$\,K is unlikely to be affected by magnetospheric processes, and therefore reflects the internal thermal evolution of the pulsar. Furthermore,  the spectrally broad thermal emission can be resolved into a function of temperature versus emitting area (c.f. Figure \ref{MW3}), which can then be used to constrain the heat conduction in the surface layers, once such models become available.
  
\subsection{Magnetospheric emission.}
The multi-wavelength data suggest that a single non-thermal component may be responsible for emission from the UV  to GeV energies. Whether this is indeed a single PL, with a high-energy cut-off, or a variable exponent would best be determined by filling in the spectral window between 10\,keV and 0.1\,GeV, which may become feasible with upcoming facilities such as {\sl NuStar}. The possible rising PL tail at low energies in Figure \ref{MW3} raises the intriguing possibility of observing the source with {\sl ALMA}. Unfortunately, MIPS on board {\sl Spitzer} is no longer operational, and the data presented here offer no constraint on the PL in the mid-IR.

The GeV luminosity observed is $8\times10^{31}$\,erg\,s$^{-1}$ (100\,MeV--100\,GeV, which contains most of the non-thermal luminosity as estimated from Figure \ref{MW3}). This gives a magnetospheric radiation efficiency $\eta=L_\gamma/\dot{E}\approx0.03$. This is within the typical range for {\sl Fermi}-detected pulsars \citep{2009ApJ...706L..56A}.

The non-thermal emission in GeV  appears to be 100\%  pulsed \citep{2010ApJS..187..460A}. In the X-ray range, the pulsed fraction increases with energy from 30\% at 0.3\,keV to over 50\% for 2--6\,keV \citep{2006ApJ...638..951Z,2007ApJ...670..668B}. This can be explained by a combination of weakly pulsed thermal component with a smooth pulse profile, and a highly-pulsed magnetospheric component with a sharper pulse profile. In detailed modeling of the energy-dependent light-curve of J0437 \citet{2007ApJ...670..668B} do not explicitly consider the role of the PL component. They do remark, however, that the pulse profile for the PL alone must not be too sharp, otherwise its effect would also be seen in the light-curves at lower energies, where thermal emission dominates. If the PL photon index were sufficiently flat ($\Gamma<2$), then the latter restriction could be lifted.

Under the assumption that the non-thermal emission is indeed a single PL component, how plausible is such a broadband PL? Assuming that we are seeing synchrotron  emission 
from FUV to $\sim$1 GeV,  a broad range of particle energies and/or magnetic field strengths must be present. There are two boundary energies: the energy of PL break ($\sim$1\,GeV) and the lowest energy observed ($\sim$1\,eV).  
With primary particles capable of radiating at up to 1\,GeV energies, a population of lower-energy secondary particles will be created, which can emit synchrotron radiation. The photon energy emitted at  the light cylinder magnetic field (where the dipolal field is the weakest), $B_{\rm LC}=3\times10^4$\,G, is approximately $E_{\rm syn}\sim10^{-4}\gamma^2$\,eV (where $\gamma$ is the Lorentz factor). Thus a lepton population near the light-cylinder with Lorentz factors in the range of a about 100 to a few times $10^6$ could account for power-law emission from optical to GeV, with the slope of the electron energy spectrum, $p=2\Gamma-1\approx2.1$. The value of the high-energy cut-off is typical for the pulsars detected by {\sl Fermi}, either classical or MSPs (e.g., \citealt{2010arXiv1012.0818B}), presumably because the $B_{LC}$ is similar for both populations. As a more conventional interpretations, the $\gamma$-ray emission could be produced by the curvature radiation or primary electrons \citep{2004ApJ...617..471M,2008ApJ...680.1378H}, with the energy of the emitted radiation dependent on the altitude of the emitting particles. However, it has been recently pointed out by \citet{2010arXiv1012.0818B} that under certain conditions, the synchrotron radiation is expected to dominate.

If the NUV emission is indeed (at least partly) magnetospheric in origin, then one would expect it to be highly pulsed, in contrast to the NS and WD thermal emission in the optical and FUV. Detection of pulsations would confirm the presence of the non-thermal component over many orders of magnitude in energy. Only the Space Telescope Imaging Spectrograph (STIS) aboard {\sl HST} has both the required sensitivity and time-resolution for such an observation. Detection of pulsation from the ground (in the $u'$-band) is possible, in principle,  but very sensitive instrumentation would be required to disentangle the magnetospheric and thermal (both WD and pulsar) emission components. Likewise, in the (long-wavelength end of) FUV, there is no current instrument but STIS capable of time resolution high enough to measure pulsation. 

Whereas non-thermal emission has been observed in the optical/NUV spectra of young pulsars such as Vela, it has not yet been found in the UV for any other old ($\tau>1$\,Gyr) pulsar. 
The late-age reappearance of a magnetospheric component in the optical may in principle result from two factors: the lower surface heating rate from the downward flux of magnetospheric particles; and that the heating of the surface from stored internal heat is expected to continue in young neutron stars only to ages of $\sim$1\,Myr. The hot thermal component is therefore much weaker in older pulsars, enabling the possible detection, as here, of both the cool thermal component from the bulk of the surface and NUV non-thermal emission.

Variable-index power-laws have been used before to describe broad-band non-thermal emission in pulsars \citep{2004AdSpR..33..495C}. Young pulsars, such as the Crab and Vela, are known to emit non-thermal emission in a very broad band. For middle-aged pulsars, extrapolation of the X-ray PL components to lower frequencies commonly over-predict the optical \citep{2007Ap&SS.308..287K}. One might speculate that for those young energetic pulsars where $B_{LC}$ is higher, the  low-energy down-turn of the PL falls between the optical and X-rays ($E_{\rm low}\propto B_{\rm LC}$), causing a break in the spectrum. On the other hand, poor modeling of the X-ray spectrum (where detailed atmospheric models may be more appropriate than simple blackbody models) may bias the non-thermal PL slopes measured. 

When the spectral fluxes of the gamma, X-ray and IR-UV data of several pulsars are plotted side-by-side, connecting the non-thermal emission with a single power-law in each case does seem plausible (see \citealt{2011arXiv1109.1984D}). That does not necessarily imply that a single component is responsible for 100\% of the non-thermal emission, but possibly for a large fraction of it. The prospect of detecting more pulsars in the optical, and improved coverage at hard X-ray energies will certainly clarify this picture in the near future.

\section{Summary and Outlook}
We have obtained broad-band spectral coverage of the binary system containing the old MSP J0437$-$4715. In particular, high-quality optical, UV and X-ray spectra, supplemented with broad-band coverage in the IR and GeV $\gamma$-rays have enabled us to construct the most complete SED for an MSP to date, as well as accurately model the atmosphere of one of the best-observed cool WDs. The data enable us to constrain the NS surface temperature and hint at a broad magnetospheric component stretching from the NUV to GeV $\gamma$-rays.

Although \psr\ gives an interesting data point in the $T(\tau)$ dependence, the thermal history of pulsars and the dependence of evolution on the mass and other parameters can only be found by measuring the surface temperatures of other pulsars through UV observations. A successful model for pulsar thermal evolution would constrain the equation of state of the neutron star interior. Similarly, an increasing sample of detailed observations of WDs with well-determined characteristics (e.g., \citealt{2011arXiv1105.3946P}) will invariably lead to better evolutionary models.

If the extremely broad non-thermal spectral component is real, it ought to be seen by {\sl nuSTAR} in the 10--50\,keV range. Similarly,  {\sl ALMA} may detect the component at lowest energies, if there is no low-energy cut-off. It will be interesting to see whether similar components reveal themselves in the spectra of other millisecond and ordinary pulsars when the full multi-wavelength spectrum is analyzed. As more {\sl Fermi} LAT data accumulates, more pulsars will be available for multi-wavelength follow-up. Multi-wavelength studies appear to be the most promising exploration tool in the coming next decade.

\bigskip\noindent
We thank M. K\"ummel for help with the reduction of prism spectral data and B. Hansen and P. Bergeron for discussion of WD spectral models.

Based on observations with {\sl HST}  (GO program 10568), VLT (program 63.H-0416(A)) and Magellan. We made use of archival data provided by the {\sl XMM-Newton} Science Archive (observation 0519\_0112320201), the {\sl Chandra} Archive (data IDs 741, 6154, 6157, 6767 and 6768), the {\sl Spitzer} Science Center (program P00023), and HEASARC for {\sl Fermi}  data.This paper includes data gathered with the 6.5 meter Magellan Telescopes located at Las Campanas Observatory, Chile. Support grant HST-GO-10568.04-A was provided by
NASA through a grant from the Space Telescope Science
Institute, which is operated by the Association of Universities
for Research in Astronomy, Inc., under NASA contract NAS5-
26555.
This work was partially supported by NASA grant NNX09AC84G and by the Ministry of Education and Science of the Russian Federation (contract 11.G34.310001).
\newline

\medskip

\bibliography{database}

\end{document}